\pdfoutput=1

\documentclass[11pt]{article}
\usepackage{epsf,amsmath,amssymb,graphicx,scalefnt,rotating,enumitem,fancyhdr}
\numberwithin{equation}{section}
\usepackage{acronym}

\usepackage[final]{showkeys}
\usepackage[noadjust]{cite}
\usepackage{filemod}
\usepackage{framed}
\usepackage{comment}
\usepackage[pdftex, colorlinks=true, linkcolor=black, filecolor=black,
  citecolor=black, urlcolor=black, draft=false, bookmarks,
  bookmarksnumbered=true, plainpages=false,
  linktocpage={true}]{hyperref}
\usepackage[affil-it]{authblk}
\usepackage[usenames,dvipsnames]{color}
\usepackage[capitalize]{cleveref}
\newcommand{\higher}{\mathrm{h.o.}}
\newcommand{\four}{four}
\newcommand{\three}{three}
\newcommand{\two}{two}
\newcommand{\one}{one}

\usepackage{slashed}
\usepackage{dsfont}

\newif\ifshownewacro
\includecomment{private}
\shownewacrotrue  
\newcounter{notecount}
\newcommand{\gfscheme}{\abbrev{GF}~scheme}

\newcommand{\lmut}{L_{\mu t}}

\newcommand{\ccf}{C_\text{F}}

\newcommand{\ctr}{T_\text{R}}
\newcommand{\nc}{n_\text{c}}
\newcommand{\na}{n_\text{A}}
\newcommand{\nf}{n_\text{f}}

\newcommand{\cca}{C_\text{A}}

\newcommand{\bare}{\text{\abbrev{B}}}
\newcommand{\ren}{\text{\abbrev{R}}}
\newcommand{\citere}[1]{Ref.\,\cite{#1}}
\newcommand{\citeres}[1]{Refs.\,\cite{#1}}
\newcommand{\code}[1]{\texttt{#1}}
\newcommand{\abbrev}[1]{{\scalefont{.9}#1}}
\newcommand{\EulerGamma}{\gamma_\text{E}}
\newcommand{\ep}{\epsilon}

\newcommand{\betagf}{\beta^\text{\abbrev{GF}}}
\newcommand{\betas}{\beta^\text{s}}
\newcommand{\betal}{\beta^\lambda}
\newcommand{\apis}{a_\text{s}}
\newcommand{\apil}{a_\lambda}
\newcommand{\apigf}{a_s^\text{\abbrev{GF}}}
\newcommand{\dd}{\mathrm{d}}
\newcommand{\deriv}[3]{\frac{\partial\ifthenelse{\equal{#1}{}}{}{^{#1}}%
    #2}{\partial #3\ifthenelse{\equal{#1}{}}{}{^{#1}}}}
\newcommand{\dderiv}[3]{\frac{\dd\ifthenelse{\equal{#1}{}}{}{^{#1}}%
    #2}{\dd #3\ifthenelse{\equal{#1}{}}{}{^{#1}}}}
\newcommand{\order}[1]{\ensuremath{{\cal O}(#1)}}

\newcommand{\msbar}{\ensuremath{\overline{\mathrm{\abbrev{MS}}}}}

\newcommand{\lhs}{l.h.s.}
\newcommand{\rhs}{r.h.s.}
\newcommand{\wrt}{w.r.t.}
\newcommand{\myacrodef}[3]{\acrodef{#2}{#3}\newcommand{#1}{\ac{#2}}}
\myacrodef{\rge}{RGE}{renormalization-group equation}
\myacrodef{\ibp}{IBP}{integration-by-parts}
\myacrodef{\qft}{QFT}{quantum field theory}
\myacrodef{\sftx}{SFTX}{short-flow-time expansion}
\myacrodef{\vev}{VEV}{vacuum expectation value}
\myacrodef{\rg}{RG}{renormalization group}
\myacrodef{\gff}{GFF}{gradient-flow formalism}
\myacrodef{\ope}{OPE}{operator product expansion}
\myacrodef{\qcd}{QCD}{quantum chromodynamics}
\myacrodef{\sqcd}{SQCD}{scalar quantum chromodynamics}
\myacrodef{\QED}{QED}{quantum electrodynamics}
\acused{QED} 
\acused{QCD} 
\myacrodef{\lhc}{LHC}{Large Hadron Collider}
\myacrodef{\uv}{UV}{ultra-violet}
\myacrodef{\lo}{LO}{leading order}
\myacrodef{\nlo}{NLO}{next-to-leading order}
\myacrodef{\nnlo}{NNLO}{next-to-next-to-leading order}
\myacrodef{\llog}{LL}{leading logarithmic}
\myacrodef{\nll}{NLL}{next-to-leading logarithmic}
\myacrodef{\nnll}{NNLL}{next-to-next-to-leading logarithmic}
\myacrodef{\pdf}{PDF}{parton density function}
\myacrodef{\sm}{SM}{Standard Model}
\myacrodef{\smeft}{SMEFT}{Standard Model Effective Field Theory}
\myacrodef{\bsm}{BSM}{beyond-the-\ac{SM}}
\myacrodef{\mssm}{MSSM}{Minimal Supersymmetric \ac{SM}}
\myacrodef{\susy}{SUSY}{Supersymmetry}
\myacrodef{\dreg}{DREG}{Dimensional Regularization}
\myacrodef{\dred}{DRED}{Dimensional Reduction}
\allowdisplaybreaks
\newcommand{\RHheaderline}{\textsf{TTK-24-057, January 2025}}
\fancypagestyle{firstpage}
{
  
  \fancyhead[R]{\RHheaderline}
}
\title{Two-loop gradient-flow renormalization of scalar QCD}
\author{Janosch Borgulat}
\author{Nils Felten}
\author{Robert V. Harlander}
\author{Jonas T. Kohnen}
\affil{TTK,
  RWTH Aachen University, 52056 Aachen, Germany}
 \date{}
\begin{document}
\maketitle
\thispagestyle{firstpage}
\begin{abstract}
  The gradient-flow formalism is applied to a non-Abelian gauge theory with
  scalar and fermionic particles, dubbed ``scalar \qcd''. It is shown that the
  flowed scalar quark requires a field renormalization, albeit only beyond
  the \one-loop level. A pseudo-physical, $t$-dependent renormalization scheme
  is defined, and the corresponding renormalization constant is evaluated at
  the \two-loop level. We also calculate the gluon action density as well as
  the condensate of the scalar and the fermionic quark at \three-loop level in
  this theory. The results validate the consistency of the gradient-flow
  formalism in this theory and provide a further step towards applying the
  gradient-flow formalism to the full Standard Model.
\end{abstract}
\parskip.0cm
\tableofcontents
\parskip.2cm

\section{Introduction}\label{sec:introduction}

The gradient flow is an established tool in lattice gauge theory
calculations~\cite{Narayanan:2006rf,Luscher:2010iy,Luscher:2009eq,
  Luscher:2011bx,Luscher:2013cpa}. It evolves the regular four-dimensional
fields into an auxiliary dimension, the flow time $t$, according to a
differential equation which is of first order in $t$. The effect of this
evolution is a suppression of the fields' high-momentum modes, and thus a
regularization of the associated \uv\ divergences.

Aside from an efficient method to determine the lattice
spacing~\cite{Luscher:2010iy,BMW:2012hcm}, or a non-perturbative definition of
the renormalized strong coupling~\cite{Luscher:2010iy}, the \gff\ also
provides a way to facilitate the calculation and renormalization of operator
matrix elements on the lattice, as well as their combination with
perturbatively evaluated Wilson coefficients. A particularly welcome feature
of the \gff\ is that it evades complications due to the breaking of Poincar\'e
invariance on the lattice. This opens efficient ways to compute matrix
elements of the energy momentum tensor on the lattice, for
example~\cite{Makino:2014taa,Suzuki:2013gza}, and most recently higher moments
of parton densities~\cite{Shindler:2023xpd,Francis:2024koo}.  A key to these
methods is the so-called \sftx, which allows one to express composite
operators of the flowed fields in terms of linear combinations of the regular
operators. Proofs of the viability of this method have been provided in
\citeres{Taniguchi:2020mgg,Suzuki:2020zue,Black:2023vju,
  Black:2024iwb,Francis:2024koo,Kim:2021qae}. It requires perturbative
matching coefficients between flowed and regular operators whose calculation
leads to non-standard Feynman diagrams and integrals. Nevertheless, the
standard methods for perturbative \qcd\ calculations can be adapted to render
such calculations possible at \nnlo\ and possibly even beyond
that~\cite{Harlander:2016vzb,Harlander:2018zpi,Artz:2019bpr,
  Harlander:2020duo,Harlander:2022tgk,Harlander:2022vgf,
  Borgulat:2023xml,Harlander:2024vmn}.

Understandably, the majority of such applications has been focused on \qcd\ up
to now. However, it is worth to consider also other theories, be it for
theoretical reasons or due to their phenomenological relevance, for example in
combination with effective field theories. Generalizations of the \gff\ have
been considered in the context of super-Yang-Mills and other supersymmetric
theories~\cite{Kikuchi:2014rla,Hieda:2017sqq,Aoki:2017iwi,
  Kasai:2018koz,Kadoh:2018qwg, Bergner:2019dim,Kadoh:2019glu,Kadoh:2022are},
or scalar theories in \three\ and \four\ space-time
dimensions~\cite{Monahan:2015lha,Aoki:2016ohw,Aoki:2016env,DelDebbio:2020amx,
  Abe:2022smm}, for example.

In this paper, we consider the extension of \qcd\ by the inclusion of scalar
quarks. While the phenomenological relevance of the literal interpretation of
this theory is doubtful in the light of current \lhc\ data, our results will
be presented for a general non-Abelian compact gauge group with a scalar field
and an arbitrary number of fermionic matter fields transforming in the
corresponding irreducible representation. They can therefore be easily
interpreted in terms of the unbroken weak sector of the \sm, for example, and
represent an important step towards a gradient-flow formulation of the full
\sm. Nevertheless, for the sake of convenience, we will refer to the
underlying theory as \sqcd\ in this paper, to the gauge bosons as gluons, and
to the fermionic and scalar matter fields as quarks and squarks (or scalar
quarks), respectively.

While the generalization of the flow equations to \sqcd\ is straightforward,
and the renormalization of the strong and the scalar coupling is the same as
in regular \sqcd\ and can be found in the literature, we will for the first
time compute the flowed field renormalizations of the quarks and the squarks
in this theory, up to \two-loop level. In addition to the \msbar\ scheme, they
will be presented in the so-called ringed scheme which is suitable also for a
lattice realization. As a first application, we will also determine the
\sftx\ of the two squark bilinear operators, namely the scalar and the Noether
current.

The remainder of this paper is structured as follows. In the next section, we
define the flow equations for \sqcd\ and provide the relevant Feynman rules.
Subsequently, the conceptual aspects for the renormalization of the strong
coupling and the flowed fields is discussed, and the relevant renormalization
constants are computed through \nnlo\ in the strong and the scalar
coupling. As an application, we study the \sftx\ of the scalar bilinear
operator and the Noether current for the scalar quarks, computing the matching
coefficients to the corresponding regular operators through \nnlo\ in
\cref{sec:sftx}. The resulting flowed anomalous dimensions of these operators
are discussed in \cref{sec:fanom}, before we present an outlook and our
conclusions in \cref{sec:conc}. In \cref{sec:renormalization}, we collect the
explicit results for the $\beta$-functions which are relevant for this paper,
and in \cref{sec:ancillary}, we describe the contents of the ancillary file
which accompanies this paper, and which provides our main results in
electronic form.

\section{Flow equations and Feynman rules}\label{sec:flow}

We consider a gauge theory with a single scalar $\phi$ (squark) and
$\nf$ fermions $\psi_f$ (quarks), all transforming in the fundamental
representation of the gauge group.  Including the gauge-fixing terms and the
corresponding Faddeev-Popov ghosts $c$, $\bar{c}$, the Lagrangian reads
\begin{equation}
  \begin{aligned}\label{eq:sqcd}
  \mathcal{L}_\text{\sqcd}&=\mathcal{L}_\text{\qcd}
  + \mathcal{L}_\phi\,,
  \\
  \mathcal{L}_\text{\qcd} &=
  -\frac{1}{4}F_{\mu\nu}^aF_{\mu\nu}^a+\sum_{f=1}^{n_f}\bar{\psi}_f
    \slashed{D}^F\psi_f
    +\frac{1}{2\xi}(\partial_\mu
    A_\mu^a)^2+\partial_\mu\bar{c}^a D_\mu^{ab}c^b\,,\\
    \mathcal{L}_\phi &=
    (D_\mu^F\phi)^\dagger(D_\mu^F\phi)
    -\frac{\lambda_\bare}{4} (\phi^\dagger\phi)^2\,.
  \end{aligned}
\end{equation}
Here, $A_\mu^a$ are the gauge fields (gluons),
\begin{equation}\label{eq:feal}
  \begin{aligned}
    D_\mu^\mathrm{F}=\partial_\mu+g_\bare A_\mu^at^a\,,\qquad
    D_\mu^{ab}=\delta^{ab}\partial_\mu-g_\bare f^{abc}A_\mu^c\,,
  \end{aligned}
\end{equation}
are the covariant derivatives in the fundamental and the adjoint
representation, respectively, and
\begin{equation}\label{eq:knot}
  \begin{aligned}
      F^{a}_{\mu\nu} = \partial_\mu A^a_\nu - \partial_\nu A^a_\mu +g_\bare
  f^{abc}A^b_\mu A^c_\nu
  \end{aligned}
\end{equation}
is the field strength tensor. The $t^a$ are the generators of the gauge group
in the fundamental representation, and the $f^{abc}$ are the structure
constants. We follow the conventions of \citeres{Luscher:2011bx,Artz:2019bpr},
which means that
\begin{equation}\label{eq:fahr}
  \begin{aligned}
    [t^a,t^b] = f^{abc}t^c\,,\qquad \mathrm{Tr}(t^at^b) = -\ctr\,\delta^{ab}\,,
  \end{aligned}
\end{equation}
with the trace normalization $\ctr$ to be specified below. It is convenient to
define the combinations
\begin{equation}\label{eq:ilka}
  \begin{aligned}
    \apis^\bare = \frac{g^2_\bare}{4\pi^2}\,,\qquad
    \apil^\bare = \frac{\lambda_\bare}{4\pi^2}\,,
  \end{aligned}
\end{equation}
where $g_\bare$ and $\lambda_\bare$ denote the bare coupling constants of the
theory.  Unless required, we will suppress the flavor indices $f$ on the quark
fields in the following.

The gradient flow evolves the fields into an auxiliary dimension, the
so-called flow time $t$. The evolution is governed by the flow equation, which
is a first-order differential equation in $t$, and associated ``initial
conditions'', which establish the contact to the regular fields at $t=0$. In
the case of the flowed gluon and quark fields $B_\mu$, $\chi$, and
$\bar{\chi}$, the flow equations are given by
\begin{equation}
 \begin{aligned}\label{eq:flow1}
  0 &= \partial_t B_\mu^a-\mathcal{D}_\nu^{ab}G^b_{\mu\nu}-\kappa
  \mathcal{D}_\mu^{ab}\partial_\nu B_\nu^b&&&\equiv \mathcal{F}_B^a\,,\\
  0 &= \partial_t \chi-\Delta\chi+g_\bare\kappa \partial_\mu B_\mu^at^a\chi
  &&&\equiv \mathcal{F}_\chi  \,,\\
  0 &= \partial_t \bar{\chi}-\bar{\chi}\overleftarrow{\Delta}-
  g_\bare\kappa \bar{\chi}\partial_\mu B_\mu^at^a
  &&&\equiv \bar{\mathcal{F}}_{\chi}
  \,,\\
 \end{aligned}
\end{equation}
and the initial conditions are
\begin{equation}\label{eq:bound}
 \begin{aligned}
  B_\mu^a(t=0,x)&=A_\mu^a(x)\,,\\
  \chi(t=0,x)&=\psi(x)\,,\\
  \bar{\chi}(t=0,x)&=\bar{\psi}(x)\,.
 \end{aligned}
\end{equation}
The gauge parameter $\kappa$ will be chosen equal to one in our calculations.
The flowed field-strength tensor is given by
\begin{align}
  G^{a}_{\mu\nu} = \partial_\mu B^a_\nu - \partial_\nu B^a_\mu +g_\bare
  f^{abc}B^b_\mu B^c_\nu\,,
\end{align}
and the covariant derivatives in the fundamental and adjoint representation
are defined as
\begin{equation}
  \begin{split}
    \mathcal{D}_\mu^\mathrm{F}=\partial_\mu+g_\bare B_\mu^at^a\,,\qquad
    \overleftarrow{\mathcal{D}}_\mu^\mathrm{F}=\overleftarrow{\partial}_\mu-g_\bare B_\mu^at^a\,,\qquad
    \mathcal{D}_\mu^{ab}=\delta^{ab}\partial_\mu-g_\bare f^{abc}B_\mu^c\,.
  \end{split}
\end{equation}
Furthermore, we introduced the short-hand notation
\begin{equation}
  \begin{split}
    \Delta = \mathcal{D}^\mathrm{F}_\mu \mathcal{D}^\mathrm{F}_\mu
    \,,\qquad
    \overleftarrow{\Delta} = \overleftarrow{\mathcal{D}}^\mathrm{F}_\mu
    \overleftarrow{\mathcal{D}}^\mathrm{F}_\mu \,.
  \end{split}
\end{equation}
For the flowed scalar fields $\varphi$, $\varphi^\dagger$, we employ the
following flow equations (see also
\citeres{Monahan:2015lha,DelDebbio:2020amx}):
\begin{equation}
 \begin{aligned}\label{eq:flow2}
   0 &= \partial_t \varphi-\Delta\varphi
   + g_\bare\kappa \partial_\mu B_\mu^at^a\varphi
   &&&\equiv\mathcal{F}_\varphi\,,\\
   0 &= \partial_t \varphi^\dagger-\varphi^\dagger\overleftarrow{\Delta}
   - g_\bare\kappa \varphi^\dagger\partial_\mu B_\mu^at^a
   &&&\equiv\mathcal{F}^\dagger_\varphi\,,
 \end{aligned}
\end{equation}
with the boundary conditions
\begin{equation}
 \begin{aligned}
  \varphi(t=0,x)&=\phi(x)\,,\\
  \varphi^\dagger(t=0,x)&=\phi^\dagger(x)\,.\\
 \end{aligned}
\end{equation}

The flow equations can be implemented in the Lagrangian by introducing
Lagrange multiplier fields $L_\mu^a$, $\lambda_f$, $\bar\lambda_f$, $\eta$,
and $\eta^\dagger$, and defining
\begin{equation}\label{eq:Ls}
  \begin{aligned}
    \mathcal{L}_{B}&=-2\int_0^{\infty} \dd t\ \text{Tr}\Big[L_\mu^a
      \mathcal{F}_B^a\Big]\,,\\
    \mathcal{L}_\chi&=\int_0^\infty \dd t
    \,\Big[\overline{\lambda}\mathcal{F}_\chi
      +\bar{\mathcal{F}}_{\chi}\lambda\Big]\,,\\
    \mathcal{L}_\varphi&=\int_0^\infty
    \dd t \,\Big[\eta^\dagger\mathcal{F}_\varphi
      + \mathcal{F}^\dagger_\varphi\eta\Big]\,,
  \end{aligned}
\end{equation}
where summation over flavor indices is understood. The flowed Lagrangian then
reads
\begin{equation}\label{eq:L}
\mathcal{L}=\mathcal{L}_\text{\sqcd}+\mathcal{L}_{B}+\mathcal{L}_{\chi}
+\mathcal{L}_{\varphi}\,,
\end{equation}
where the first term notes the regular (i.e., unflowed) part of the Lagrangian
given in \cref{eq:sqcd}, and the remaining terms incorporate the flow
equations in \cref{eq:flow1,eq:flow2}.

Using \cref{eq:L,eq:Ls}, one can derive Feynman rules for the flowed fields
along the lines of \citere{Luscher:2011bx}. In addition to generalized
propagators for the flowed fields, they also involve so-called \textit{flow
  lines}, corresponding to mixed two-point functions of the flowed field and
its associated Lagrange multiplier. They couple to one another or to flowed
fields via \textit{flowed vertices} which involve integrations over flow-time
parameters.  For details, we refer to \citere{Artz:2019bpr}, where also the
complete set of Feynman rules for the non-scalar part of \cref{eq:L} can be
found. For the regular \sqcd\ Feynman rules involving scalars, on the other
hand, we use\footnote{All diagrams in this paper are produced with the help of
\textsc{FeynGame}~\cite{Harlander:2020cyh,Harlander:2024qbn,Bundgen:2025utt}.}
\begin{equation}\label{eq:dona}
  \begin{aligned} \raisebox{-3.6em}{\includegraphics[%
    width=.22\textwidth]{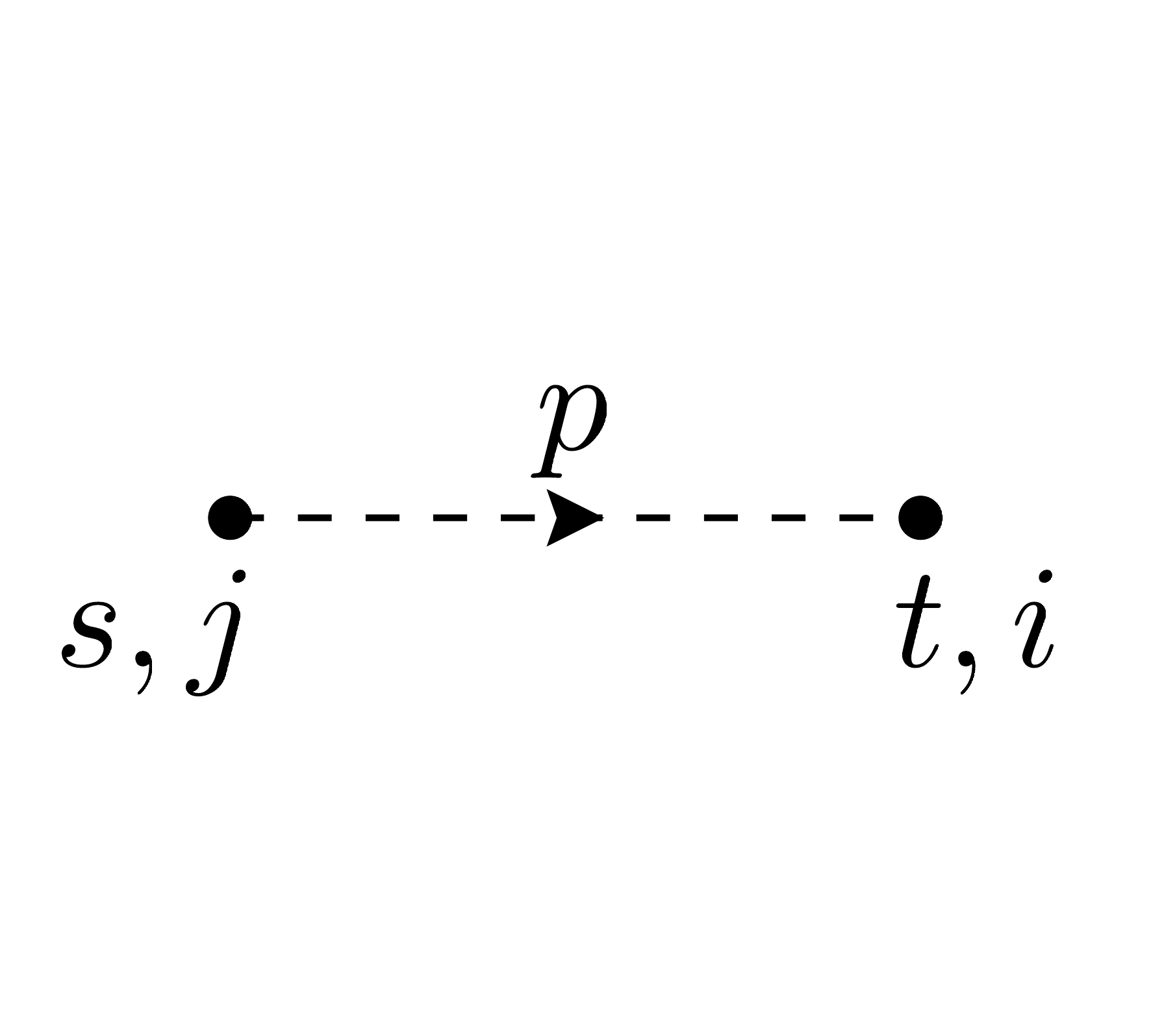}} &
    =\delta_{ij}\frac{1}{p^2}e^{-(t+s)p^2}\,,\\ \raisebox{-3.6em}{\includegraphics[%
    width=.22\textwidth]{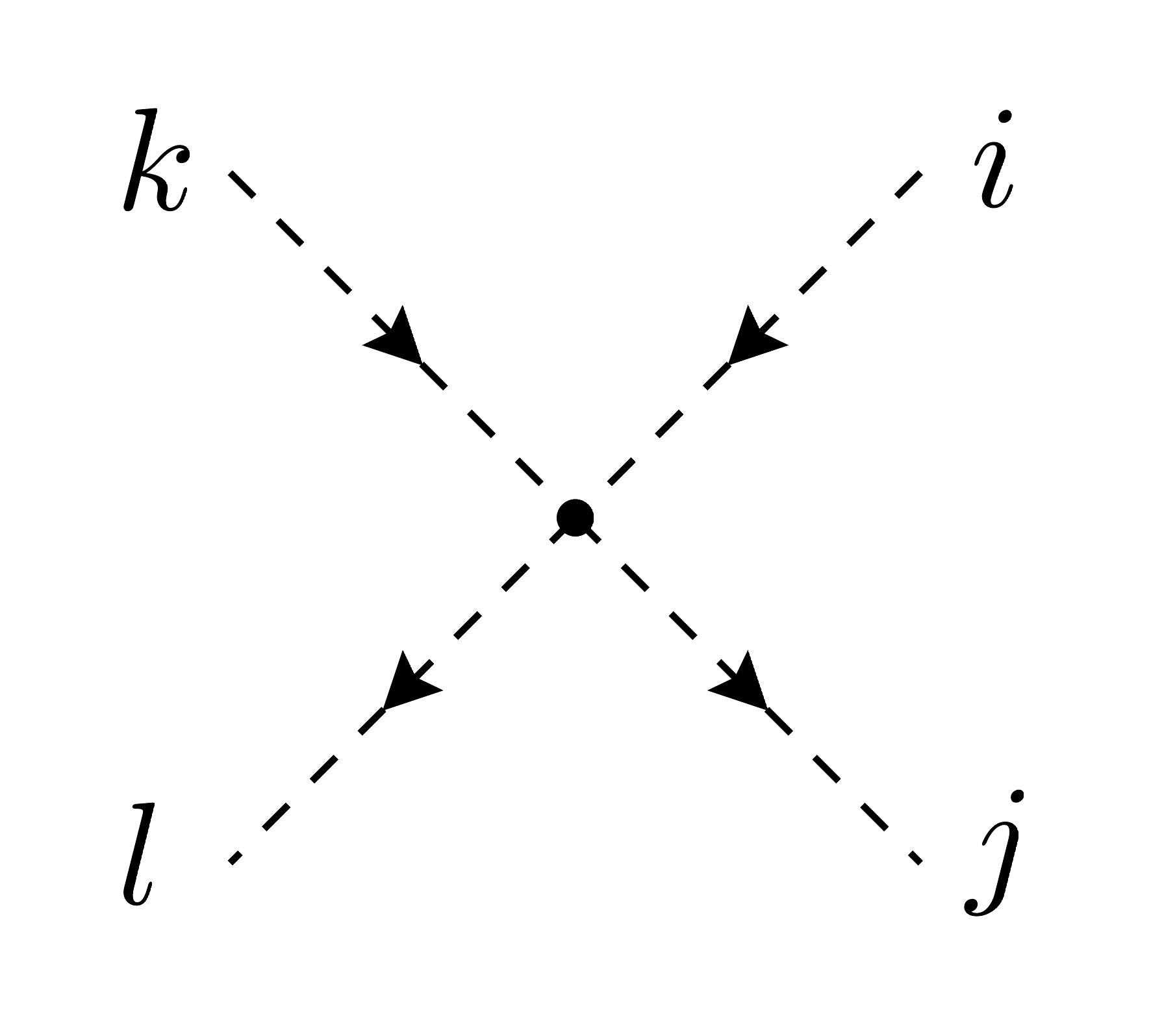}} &
    =-\frac{\lambda^\bare}{2} \left(\delta_{ij}\delta_{kl}
    +\delta_{il}\delta_{jk}\right)\,,\\ \raisebox{-3.6em}{\includegraphics[%
    width=.22\textwidth]{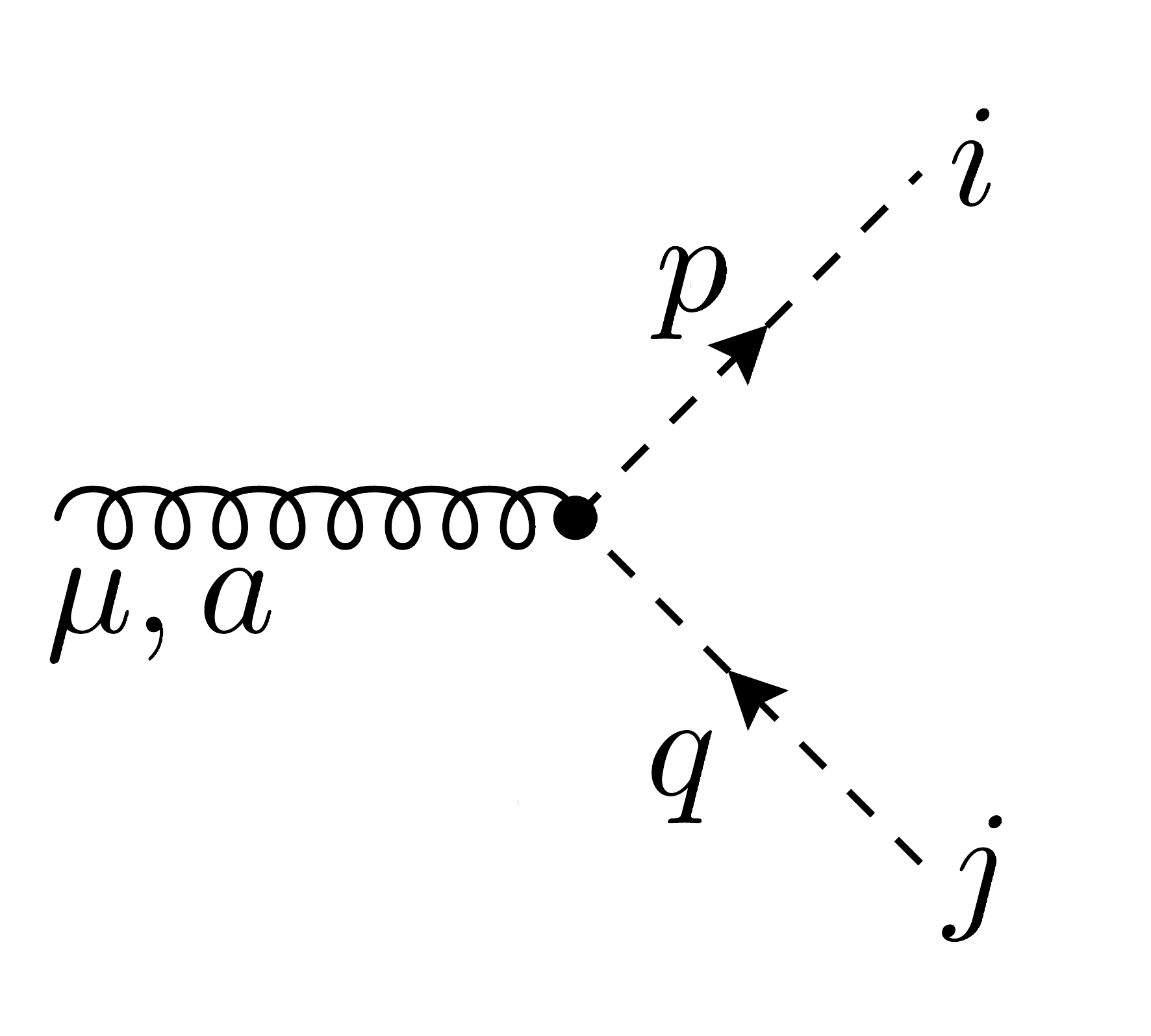}} &
    =ig_\bare(t^a)_{ij}\ \left(p^\mu-q^\mu\right)\,,\\ \raisebox{-3.6em}{\includegraphics[%
    width=.22\textwidth]{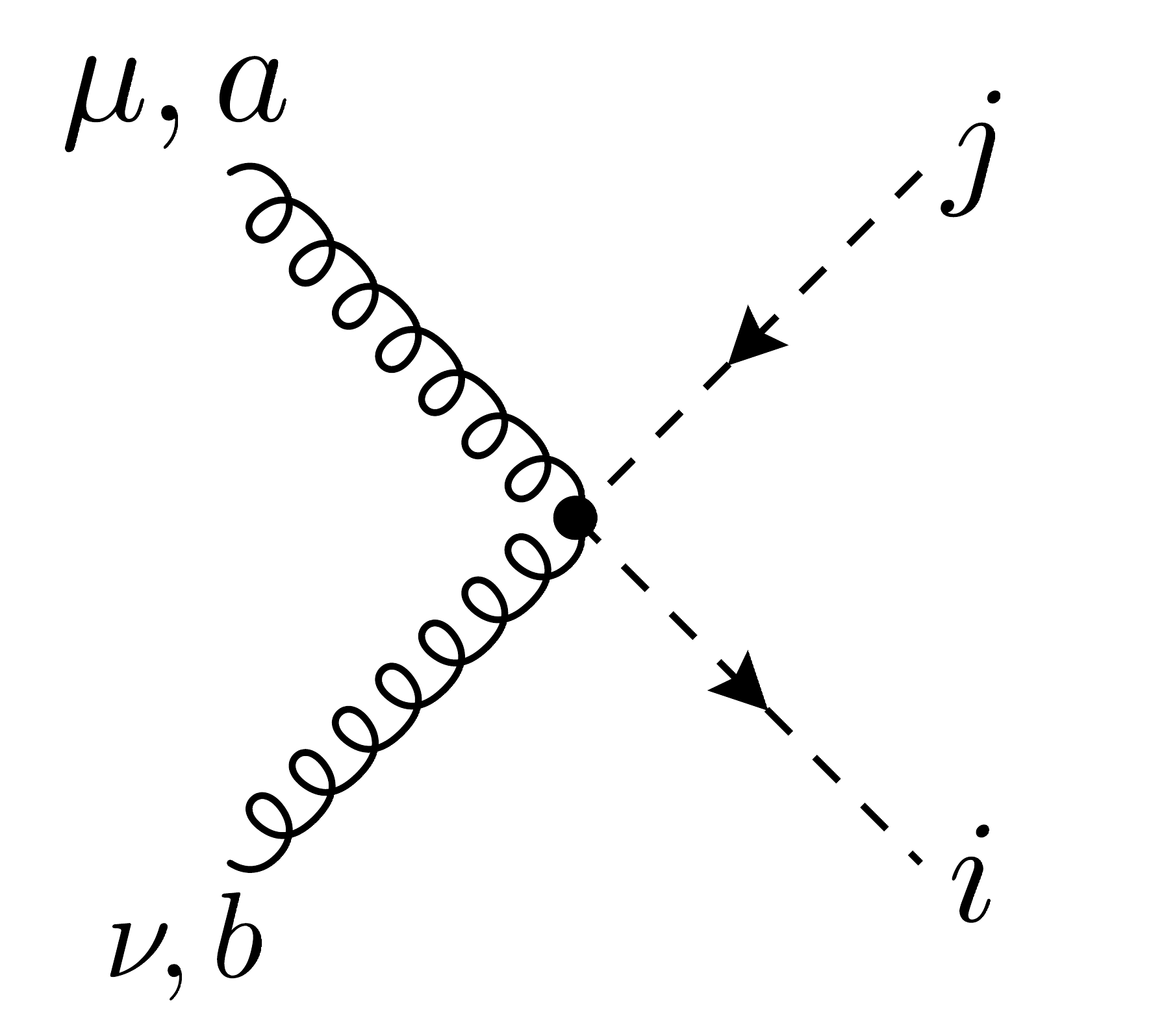}} &
    =g^2_\bare\{t^a,t^b\}_{ij}\ \delta_{\mu\nu}\,.  \end{aligned}
\end{equation}
Here, $i,j,k,l$ are color indices of the fundamental representation, $a,b$ are
color indices of the adjoint representation, $\mu,\nu$ are Lorentz indices,
$s$ and $t$ are flow-time variables, and the momenta $p$ and $q$ are assumed
outgoing. The Feynman rules for the flowed vertices read
\begin{equation}
  \begin{aligned}
    \raisebox{-3.6em}{\includegraphics[%
        width=.22\textwidth]{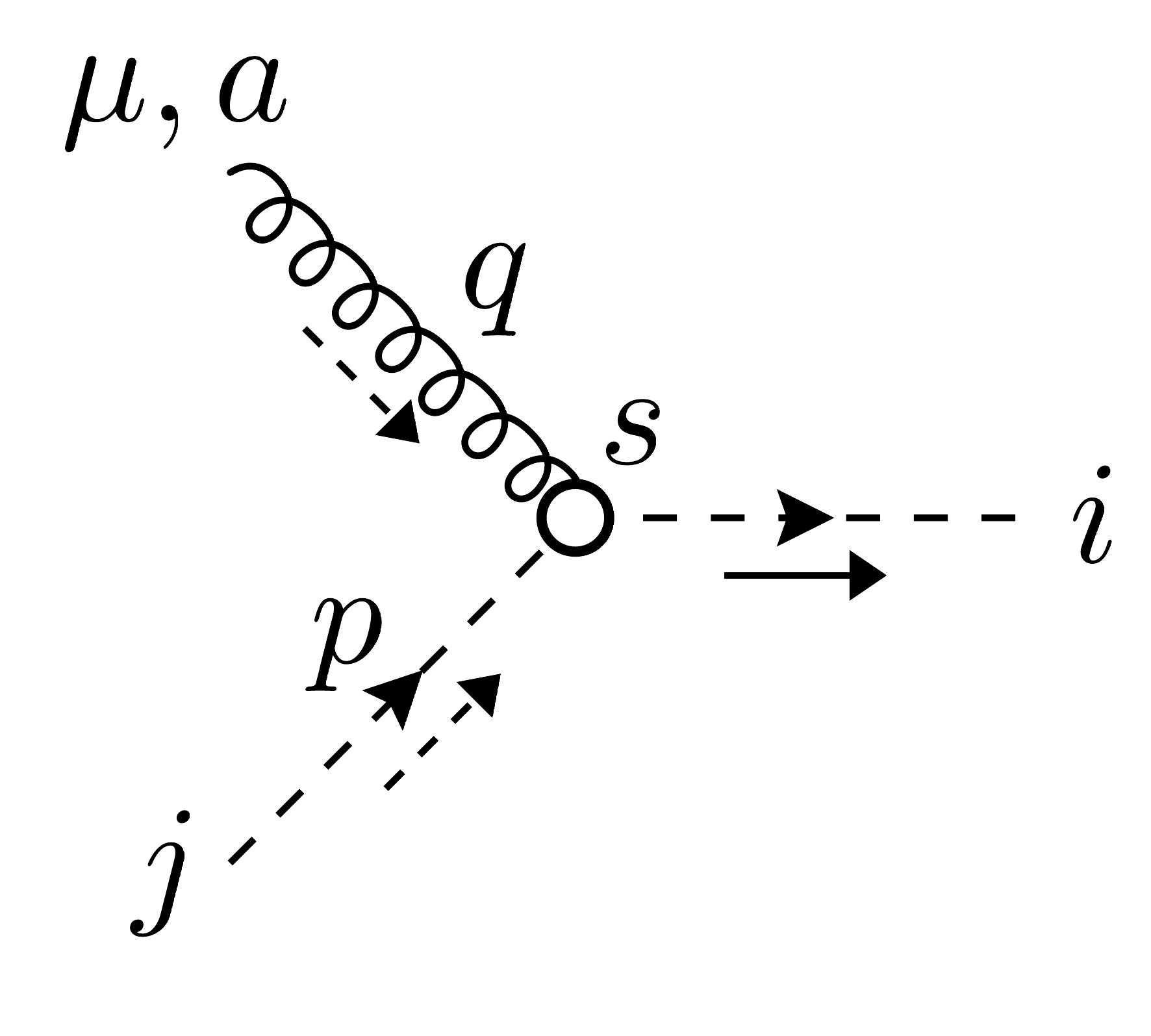}}
    &=ig_\bare (t^a)_{ij}\int_0^\infty ds\left(2p_\mu+(1-\kappa)q_\mu\right)\,,\\
    \raisebox{-3.6em}{\includegraphics[%
        width=.22\textwidth]{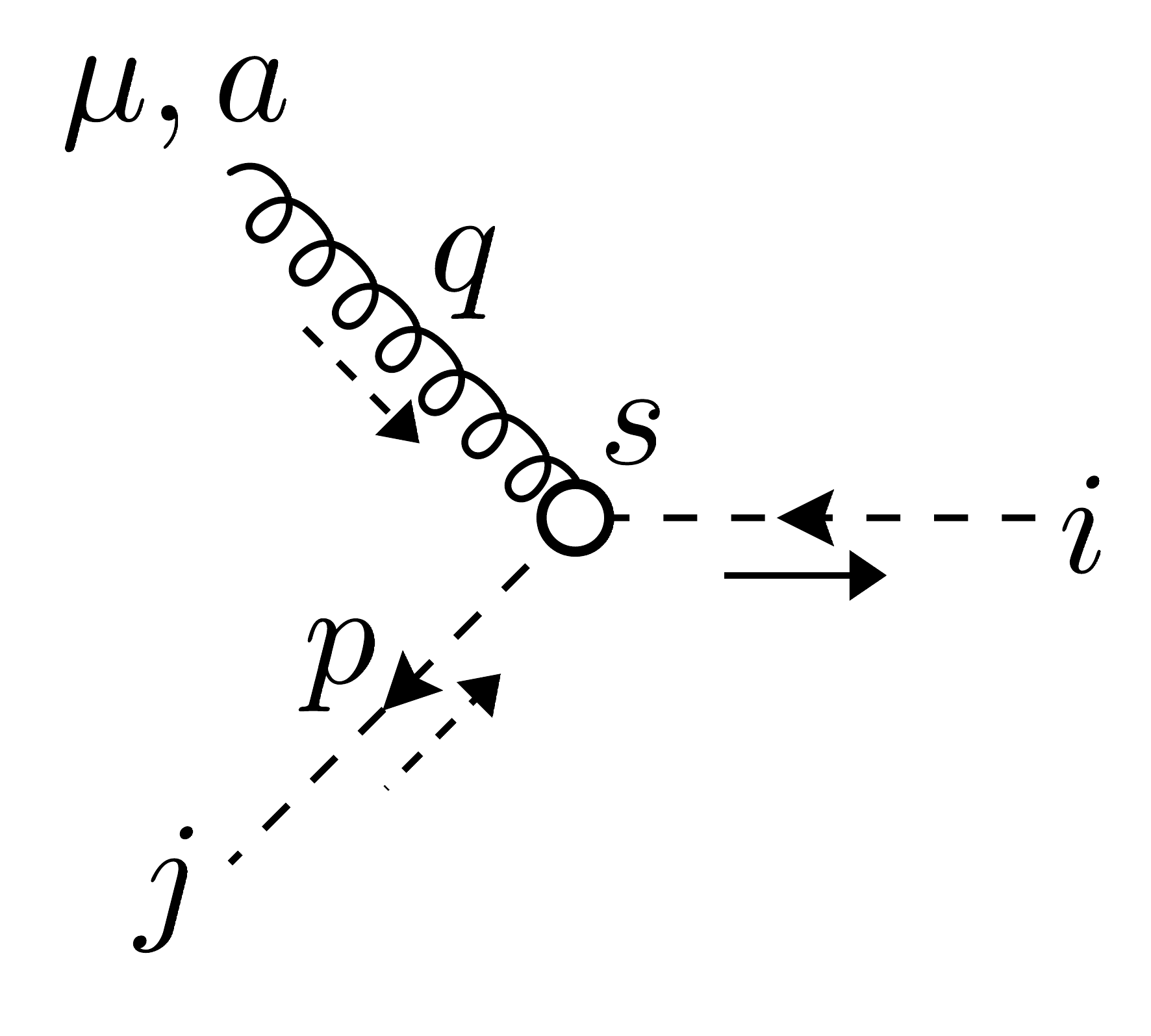}}
    &=-ig_\bare (t^a)_{ji}\int_0^\infty ds\left(2p_\mu+(1-\kappa)q_\mu\right)\,,\\
    \raisebox{-3.6em}{\includegraphics[%
        width=.22\textwidth]{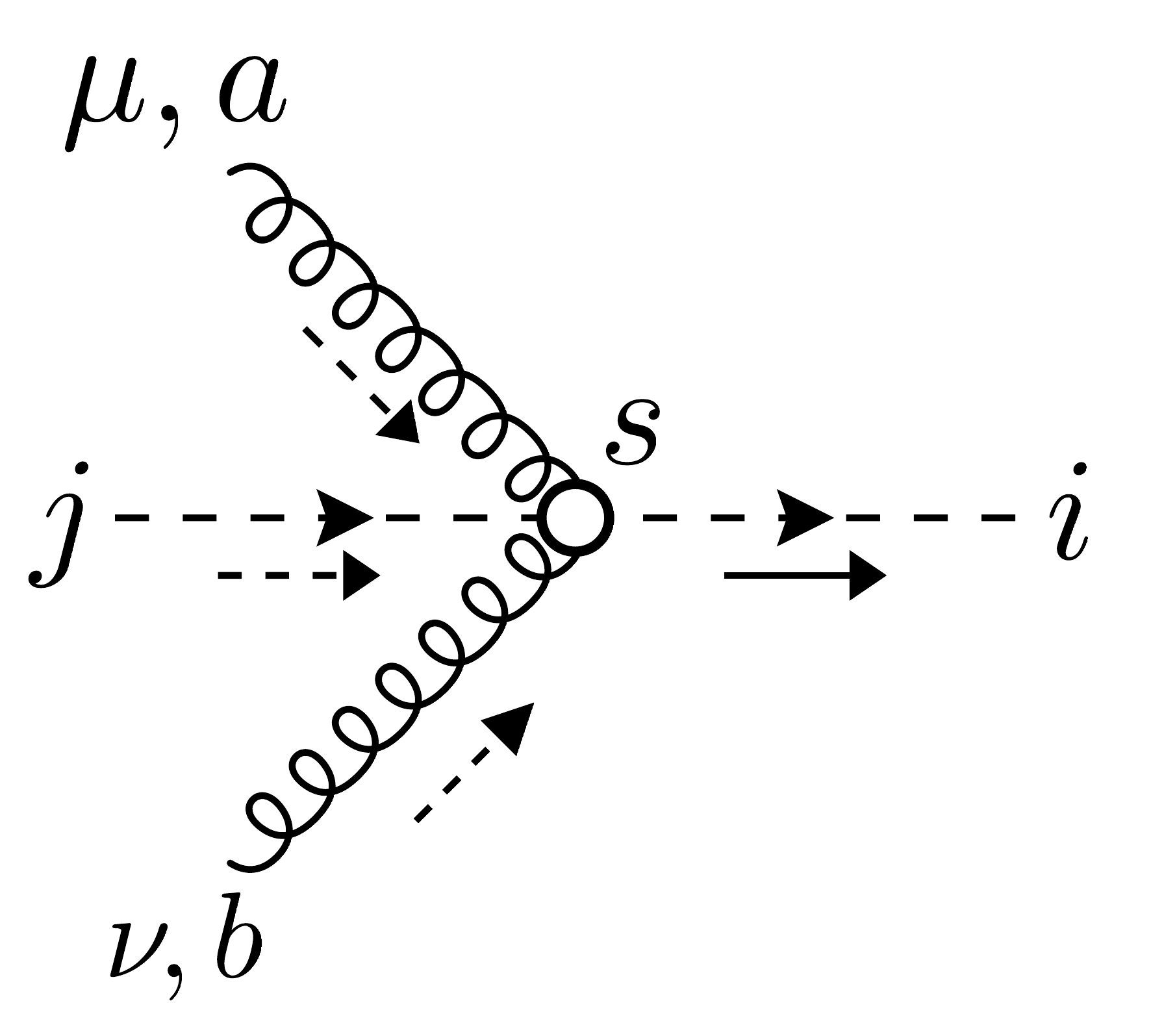}}
    &=g^2_\bare\{t^a,t^b\}_{ij}\ \delta_{\mu\nu}\int_0^\infty ds\,,\\
    \raisebox{-3.6em}{\includegraphics[%
        width=.22\textwidth]{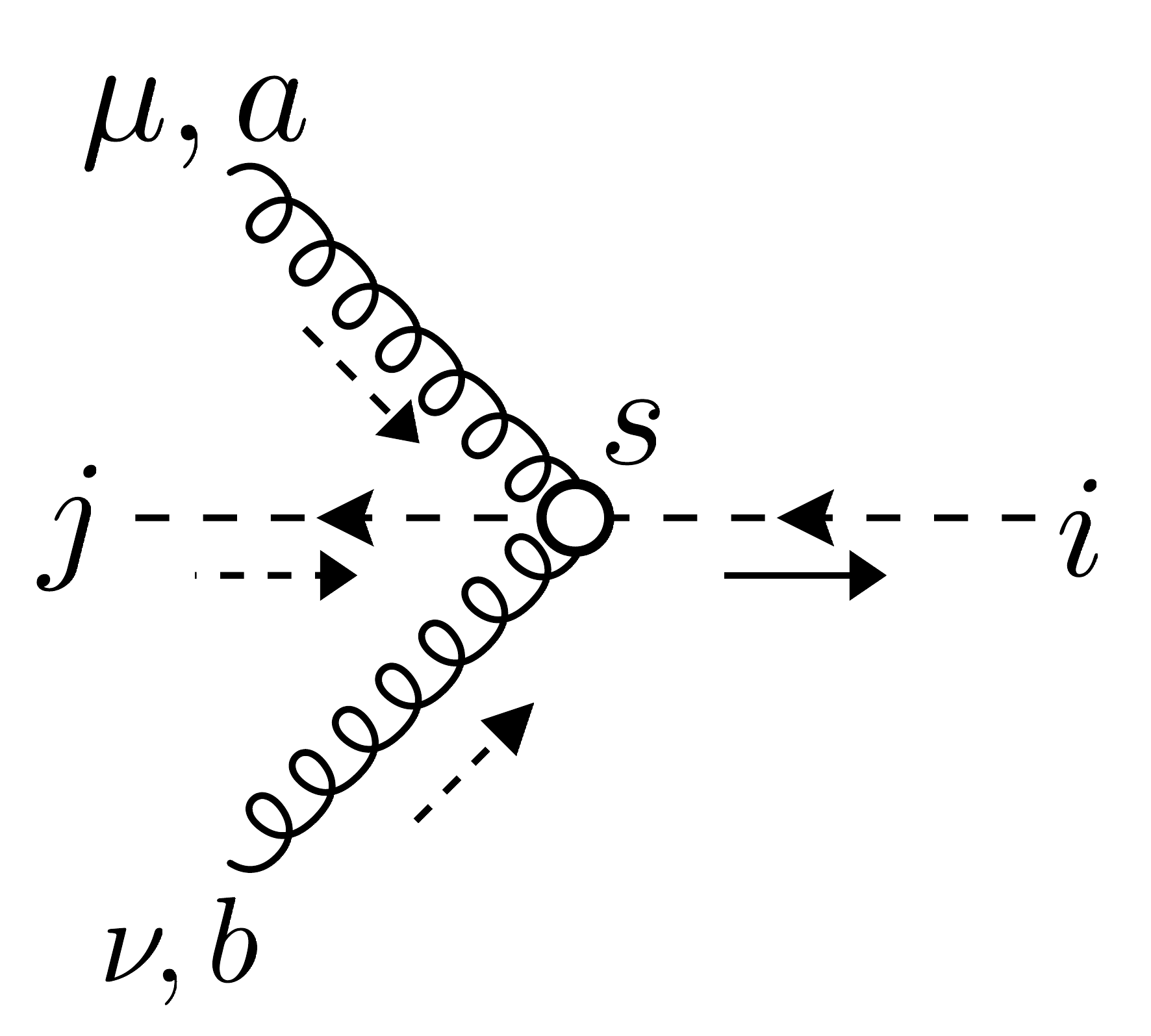}}
    &=g^2_\bare\{t^a,t^b\}_{ij}\ \delta_{\mu\nu}\int_0^\infty ds\,,
  \end{aligned}
\end{equation}
where the arrows next to the lines, indicating that the corresponding line is
a flow line, point towards increasing flow time. Dashed arrows indicate that
the line can be a flowed propagator or a flow line.

\section{Renormalization constants}\label{sec:ren}

It is well known that the renormalization of the fundamental parameters of the
flowed Lagrangian is the same as for the regular
theory~\cite{Luscher:2011bx}. Since we assume the quarks and the squark to be
massless, the only fundamental parameters of our theory are the coupling
constants $\apis^\bare$ and $\apil^\bare$. Their renormalization in \sqcd\ is
known through \three-loop level~\cite{Steudtner:2024teg} in the
\msbar\ scheme, which, however, is only defined perturbatively. The \gff, on
the other hand, allows for a definition of the strong coupling which can be
implemented both perturbatively and non-perturbatively:
\begin{equation}\label{eq:harp}
  \begin{aligned}
    \frac{\alpha_s^\text{\abbrev{GF}}(t)}{\pi} \equiv \apigf(t)
    &= \frac{E(t)}{E_0(t)}\,,\quad\text{with}\quad
    E_0(t) = \frac{3\na}{32 t^2}\,,
  \end{aligned}
\end{equation}
where
\begin{equation}\label{eq:broz}
  \begin{aligned}
    E(t) &= \frac{g_\bare^2}{4}\langle G_{\mu\nu}^a(t)G_{\mu\nu}^a(t)\rangle
  \end{aligned}
\end{equation}
is the gluon action density, and $\na$ is the dimension of the adjoint
representation of the gauge group.  The perturbative expression for $E(t)$
within \qcd, and thus the conversion between the \msbar\ and the \gfscheme, is
currently known through \nnlo~\cite{Harlander:2016vzb} and has been used for a
novel extraction of the \qcd\ scale $\Lambda_\text{\qcd}$ in the
\msbar\ scheme~\cite{Hasenfratz:2023bok} and the \msbar\ coupling
$\alpha_s(M_Z)$~\cite{Wong:2023jvr}. In \cref{sec:action}, we will generalize
this result to \sqcd. A similar non-perturbative definition is also possible
for the scalar coupling $\alpha_\lambda$, of course, but it requires the
calculation of $\langle(\phi^\dagger\phi)^2\rangle$, which is beyond the scope
of the current paper.

Concerning the renormalization of the flowed fields, we write
\begin{equation}\label{eq:cuif}
  \begin{aligned}
    B^{a,\ren}_\mu&=Z_B^{1/2}B_\mu^a\,,&&\\
    \chi_f^\ren&=Z_\chi^{1/2}\chi_f\,,&
    \bar\chi_f^\ren&=Z_\chi^{1/2}\bar\chi_f\,,
    \\
    \varphi^\ren&=Z_\varphi^{1/2}\varphi\,,&
    \varphi^{\dagger,\ren}&=Z_\varphi^{1/2}\varphi^\dagger\,.
  \end{aligned}
\end{equation}
It was shown in \qcd\ that the flowed gluon does not require
renormalization~\cite{Luscher:2011bx,Luscher:2013cpa}, and since the same
arguments apply to \sqcd, we can set $Z_B=1$ to all orders in perturbation
theory.

On the other hand, the flowed quark field turns out to require a non-trivial
renormalization $Z_\chi$. A particularly useful renormalization condition
which can be implemented both on the lattice and in perturbation theory is the
so-called ringed scheme, defined as
\begin{equation}\label{eq:dawn}
  \begin{aligned}
    \mathring{Z}_\chi\sum_{f=1}^{\nf} \langle\bar\chi_f(t)
    \slashed{\mathcal{D}}^\mathrm{F}\!  \chi_f(t)\rangle =
    -\frac{2\nc\nf}{(4\pi t)^2}\,,
  \end{aligned}
\end{equation}
where $\nc$ is the dimension of the fundamental representation of the gauge
group. 
In \qcd, $\mathring{Z}_\chi$ is known through
\nnlo~\cite{Luscher:2013cpa,Artz:2019bpr}. In this paper, we will generalize
this result to \sqcd, see \cref{sec:fermion}.

The flowed squark renormalization $Z_\varphi$ has not been considered
before. Its \nnlo\ expression will be one of the main results of our
calculation. We provide it in a scheme analogous to the fermionic case by
requiring the renormalization condition
\begin{equation}\label{eq:hark}
  \begin{aligned}
    \mathring{Z}_\varphi
    \langle\varphi^\dagger(t)\varphi(t)\rangle
    \equiv \frac{\nc}{32\pi^2 t}\,,
  \end{aligned}
\end{equation}
where the \rhs\ is defined such that $\mathring{Z}_\varphi=1+\text{higher
orders}$.

\subsection{The strong coupling of scalar \abbrev{QCD} in the
  gradient-flow scheme}\label{sec:action}

\begin{figure}
  \begin{center}
    \begin{tabular}{ccc}
    \hspace*{-1em}
    \includegraphics[%
            clip,width=.3\textwidth]%
                          {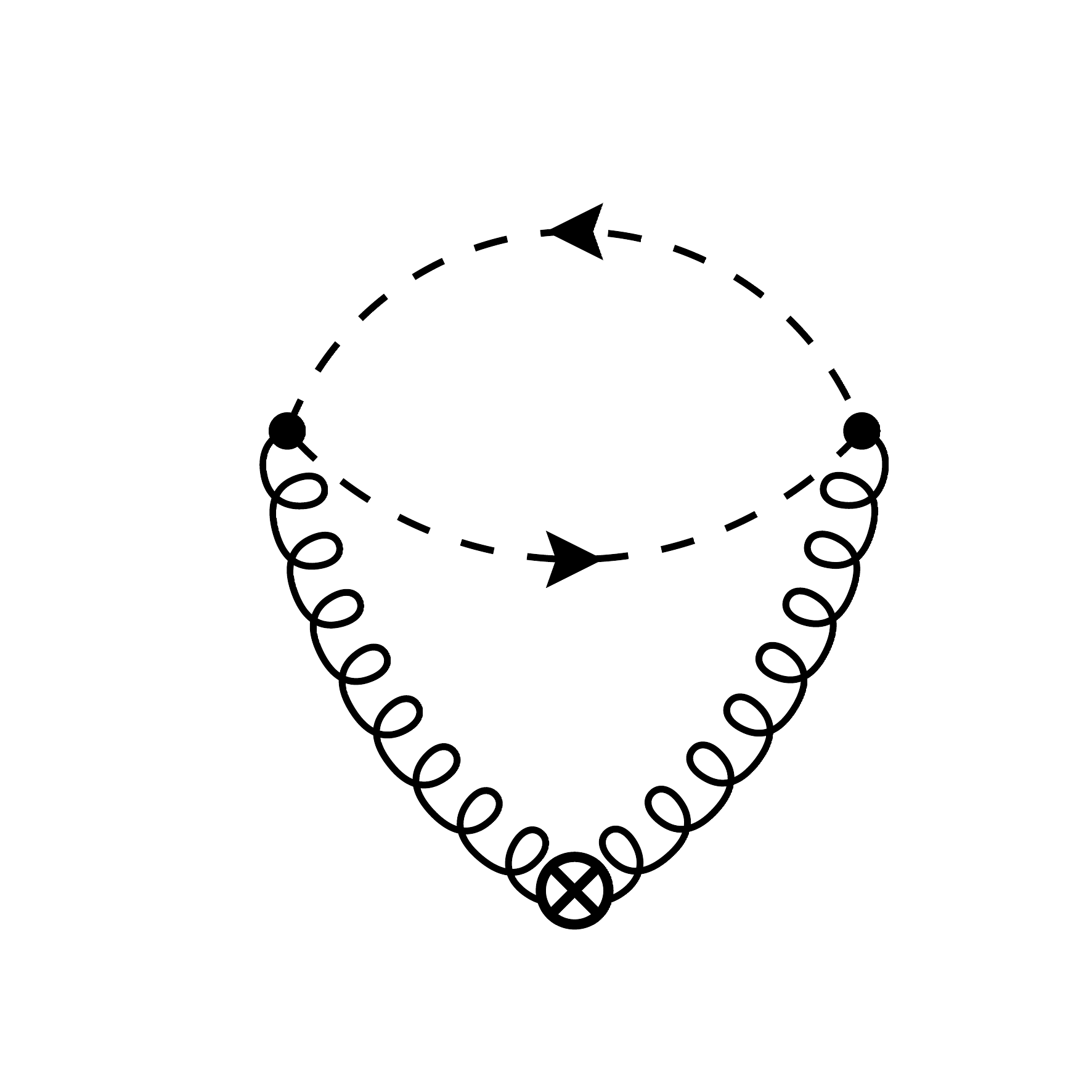}
                          &
    \hspace*{-1em}
          \includegraphics[%
            clip,width=.3\textwidth]%
                          {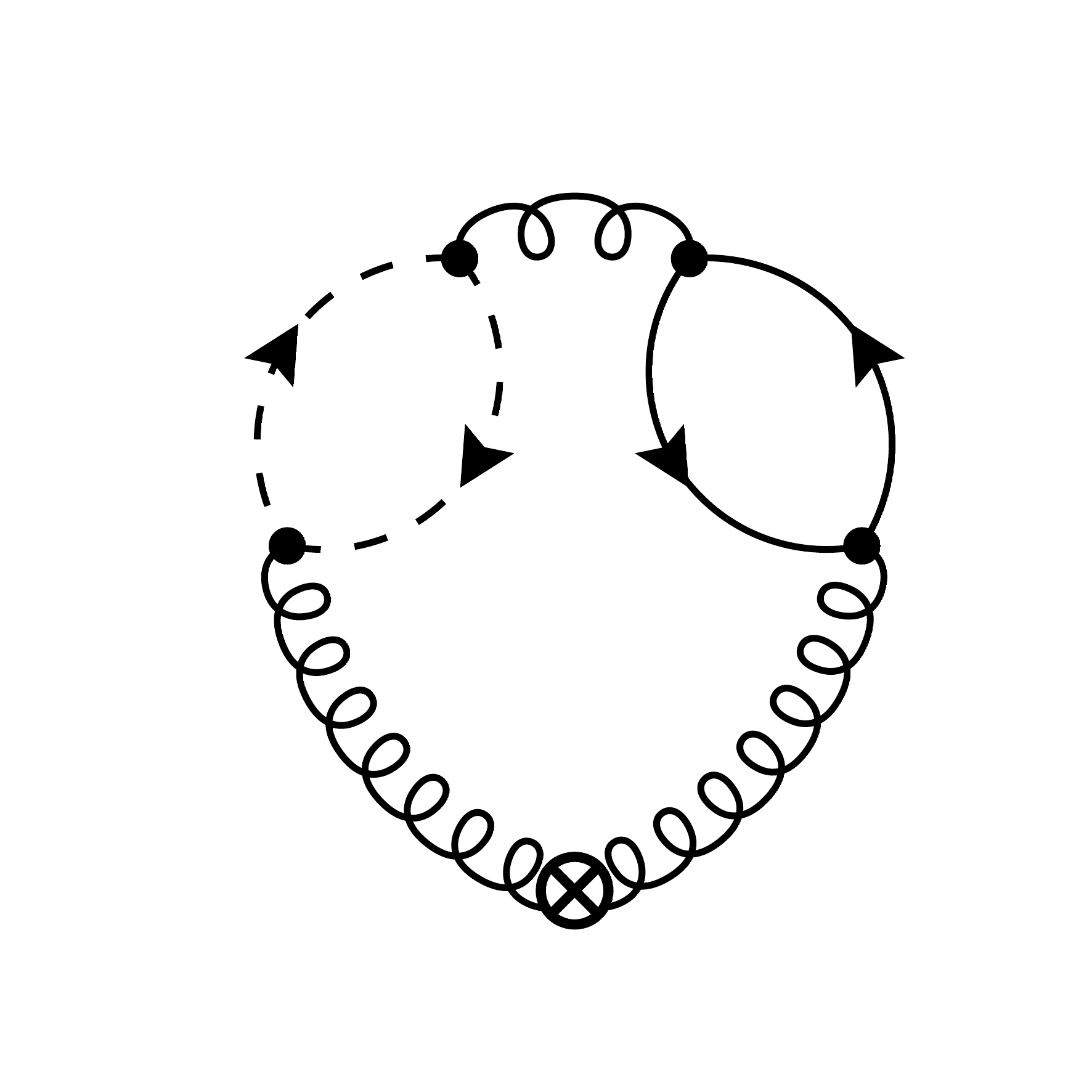}
                          &
    \hspace*{-1em}
          \includegraphics[%
            clip,width=.3\textwidth]%
                          {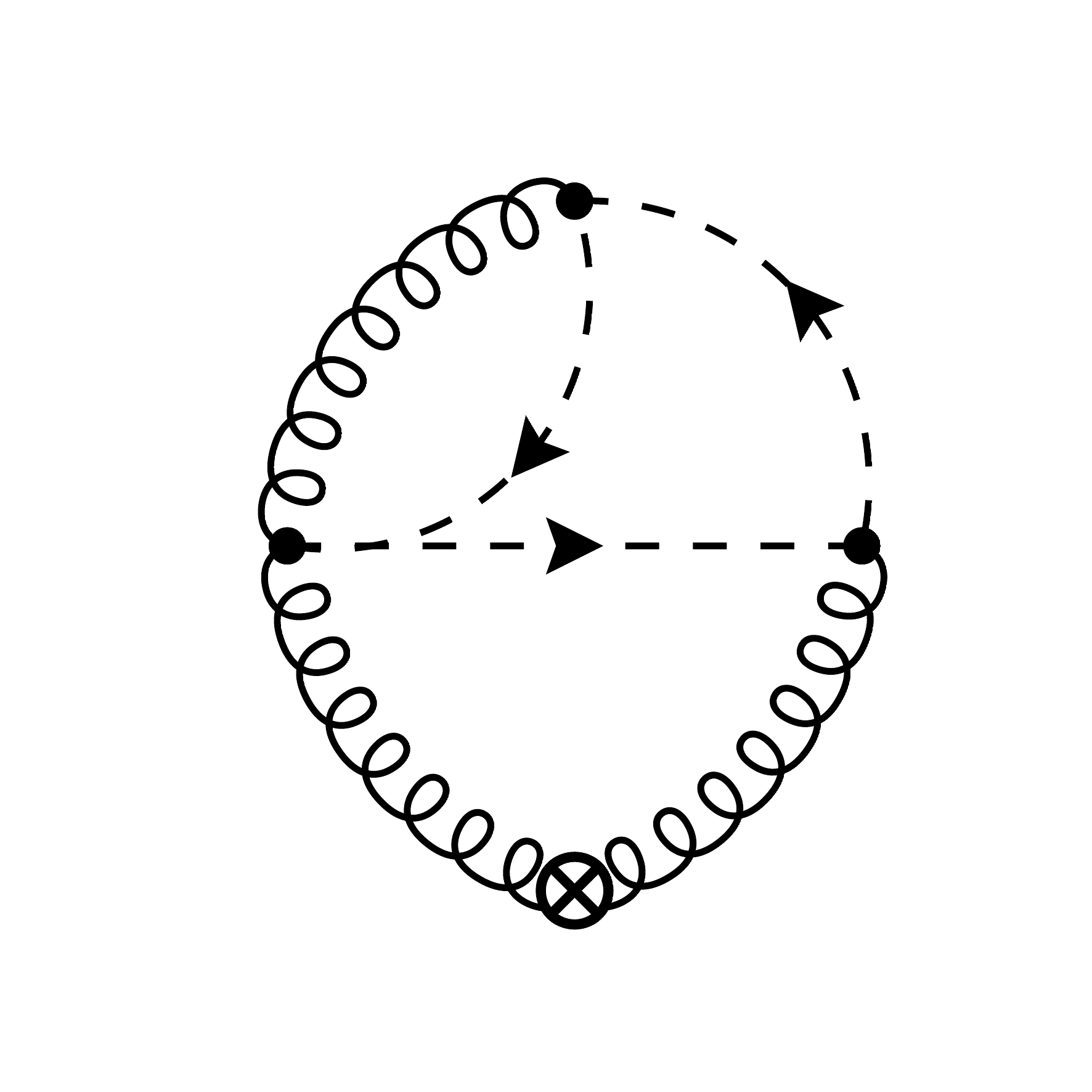}\\
                          (a) & (b) & (c)
    \end{tabular}
    \parbox{.9\textwidth}{
      \caption[]{\label{fig:GGdias}\sloppy Contributions to $E(t)$ involving
        squarks at (a)~\nlo{} and (b), (c)~\nnlo. The spiral lines are gluons,
        the dashed lines are squarks, and the solid lines are quarks. The
        lower vertex marked $\mathbf{\otimes}$ denotes the flowed operator
        $G_{\mu\nu}(t)G^{\mu\nu}(t)$, the other vertices are regular
        \qcd\ vertices.  }}
  \end{center}
\end{figure}

In order to generalize the result for \cref{eq:broz} to \sqcd, we need to take
into account diagrams with closed squark loops. The only such diagram which
gives a non-zero contribution up to the \two-loop level is shown in
\cref{fig:GGdias}\,(a); all others which could be generated from the Feynman
rules vanish because they lead to scaleless loop integrals. Two \three-loop
diagrams with squark loops are shown in \cref{fig:GGdias}\,(b),(c). Since the
associated loop integrals are \uv\ divergent before the renormalization of
$\alpha_s$, we adopt dimensional regularization and evaluate them in
$D=4-2\ep$ space-time dimensions.

All calculations in this paper are carried out by using the tool chain first
described in \citere{Artz:2019bpr}, albeit slightly updated. It is based on
\code{qgraf}~\cite{Nogueira:1991ex,Nogueira:2006pq} for the generation of the
Feynman diagrams, \code{tapir}~\cite{Gerlach:2022qnc} for inserting the
Feynman rules, \code{exp}~\cite{Harlander:1998cmq} for the identification of
the integral topologies, \code{FORM}~\cite{Vermaseren:2000nd,Kuipers:2012rf}
for all algebraic operations, and
\code{Kira}+\code{FireFly}~\cite{Maierhofer:2017gsa,
  Maierhofer:2018gpa,Klappert:2019emp,Klappert:2020nbg,Klappert:2020aqs} for
the reduction to master integrals.  At \two-loop level, all master integrals
are known in analytic form~\cite{Luscher:2010iy,Harlander:2018zpi}.  At
\three-loop level, we evaluate the master integrals using
\code{ftint}~\cite{Harlander:2024vmn}, which uses
\code{pySecDec}~\cite{Borowka:2015mxa,Borowka:2017idc,
  Borowka:2018goh,Heinrich:2023til} to perform a sector
decomposition~\cite{Binoth:2000ps,Binoth:2003ak} and to evaluate the
coefficients of the Laurent series in $\ep$ numerically.

After renormalization of the strong coupling according to \cref{eq:coupdef},
we obtain the result
\begin{equation}\label{eq:knag}
  \begin{aligned}
    E(t) = \apis E_0(t)\,
    \left(1 + e_1 + e_2 + \higher\right)
  \end{aligned}
\end{equation}
with
\begin{equation}\label{eq:funk}
  \begin{aligned}
    e_n &= \sum_{i=0}^n \left[e_{i,n-i} + e^{(1l)}_{i,n-i}\lmut
      + \cdots + e_{i,n-i}^{(nl)}\lmut^n\right]
    \apis^i\apil^{n-i}\,,
  \end{aligned}
\end{equation}
and
\begin{equation}\label{eq:abed}
  \begin{aligned}
    \lmut = \ln 2\mu^2 t + \EulerGamma\equiv \ln\frac{\mu^2}{\mu_t^2}\,,
  \end{aligned}
\end{equation}
where $\EulerGamma = -\Gamma'(1)=0.577216\ldots$ is the Euler-Mascheroni
constant, and we have implicitly defined the $t$-dependent energy scale
$\mu_t$.  Here and in the following, $\apis$ and $\apil$ denote the
\msbar\ renormalized couplings defined in \cref{eq:coupdef}, with their
dependence on the renormalization scale $\mu$ suppressed, and ``$\higher$''
(for ``higher orders'') denotes terms of order $\apis^i\apil^j$, with $i+j\geq
3$.  For the constant terms up to \three-loop level ($n\leq 2$)
in \cref{eq:funk}, we obtain
\begin{equation}\label{eq:funk1}
  \begin{aligned}
    e_{10} &= e_{10}^\text{\qcd} - \frac{5}{36}\ctr\,,
    \\
    e_{20} &= e_{20}^\text{\qcd} - \ctr\left[0.7639\,\cca
      + 0.34532\,\ccf - \ctr\left(0.0976\,\nf + 0.0238\right)\right]\,,\\
    e_{ij} &= 0\qquad\text{otherwise}\,,
  \end{aligned}
\end{equation}
where the color factors are
\begin{equation}\label{eq:acti:kahn}
  \begin{aligned}
    \ccf &= \ctr\frac{\nc^2-1}{\nc}\,,\qquad
    \cca = 2\,\ctr\nc\,,\qquad \ctr = \frac{1}{2}\,.
  \end{aligned}
\end{equation}

The coefficients $e_{10}^\text{\qcd}$ and $e_{20}^\text{\qcd}$
correspond to the pure \qcd\ result as obtained in
\citere{Harlander:2016vzb,Artz:2019bpr}. Since we adopt a different
normalization in this paper, let us quote them again here:
\begin{equation}\label{eq:epis}
  \begin{aligned}
    e_{10}^{\text{\qcd}} &=
    \left(\frac{13}{9} + \frac{11}{6} \ln 2 - \frac{3}{4} \ln 3\right)
    \cca - \frac{2}{9} \ctr\nf \\
    e_{20}^{\text{\qcd}} &= 1.74865\, \cca^2  - (1.97283\ldots)\, \ctr\cca\nf
    \\&\quad+ \left(\zeta_3 - \frac{43}{48}\right)
    \ctr\ccf\nf + \left(\frac{1}{9}\zeta_2 -
    \frac{5}{81}\right) \ctr^2\nf^2\,,
  \end{aligned}
\end{equation}
where $\zeta_n\equiv \sum_{k=1}1/k^n$ is Riemann's zeta function with
$\zeta_2=\pi^2/6=1.64493\ldots$ and $\zeta_3=1.20206\ldots$.  The exact
expression for the $\ctr\cca\nf$ term can be found in \citere{Artz:2019bpr}.

The logarithmic terms of \cref{eq:funk} can be derived from the \rge
\begin{equation}\label{eq:alow}
  \begin{aligned}
    \mu^2\dderiv{}{}{\mu^2}E(t) &= \left[\deriv{}{}{\lmut} +
      \betas(\apis,\apil)\deriv{}{}{\apis}
      + \betal(\apis,\apil)\deriv{}{}{\apil}\right]E(t) = 0\,,
  \end{aligned}
\end{equation}
with the beta functions $\betas$ and $\betal$ defined in
\cref{eq:betadef}. For $n\leq 2$, this leads to
\begin{equation}\label{eq:funk2}
  \begin{aligned} e_{10}^{(1l)} &= \betas_{20}\,,\qquad e^{(1l)}_{20}
    = \betas_{30} + 2\,\betas_{20}\, e_{10}\,,\qquad e^{(2l)}_{20} =
    (\betas_{20})^2\,,\\ e_{ij}^{(nl)} &=
    0\qquad\text{otherwise}\,.  \end{aligned}
\end{equation}
The coefficients $\betas_{ij}$ are collected in \cref{eq:ceyx}.

Since $E(t)$ and the strong beta function in the \msbar\ scheme are
independent of $\apil$ through \three-loop level, our results allow us to
evaluate the beta function of the strong coupling in \sqcd\ in the
gradient-flow scheme. In analogy to \cref{eq:betadef}, we write
\begin{equation}\label{eq:amah}
  \begin{aligned}
    \mu^2\frac{d}{d\mu^2}\apigf(\mu)&=\betagf(\apigf,\apil)
    \equiv-\ep\,\apigf
    -\sum_{n=0}^{3} \betagf_{n0}
    (\apigf)^n + \higher
  \end{aligned}
\end{equation}
The first two perturbative coefficients are scheme independent, of course,
thus
\begin{equation}\label{eq:alva}
  \begin{aligned}
    \betagf_{20} &= \betas_{20}\,,\qquad
    \betagf_{30} = \betas_{30}\,,
  \end{aligned}
\end{equation}
while the \three-loop coefficient is given by
\begin{equation}\label{eq:coda}
  \begin{aligned}
    \betagf_{40} &= \betas_{40} - e_{10}\betas_{30} + \left(e_{20} -
    e_{10}^2\right)\betas_{20}\,.
  \end{aligned}
\end{equation}

\subsection{The flowed fermionic field renormalization}\label{sec:fermion}

\begin{figure}
  \begin{center}
  \begin{tabular}{cc}
    \hspace*{.2em}
    \includegraphics[%
    clip,width=.3\textwidth]{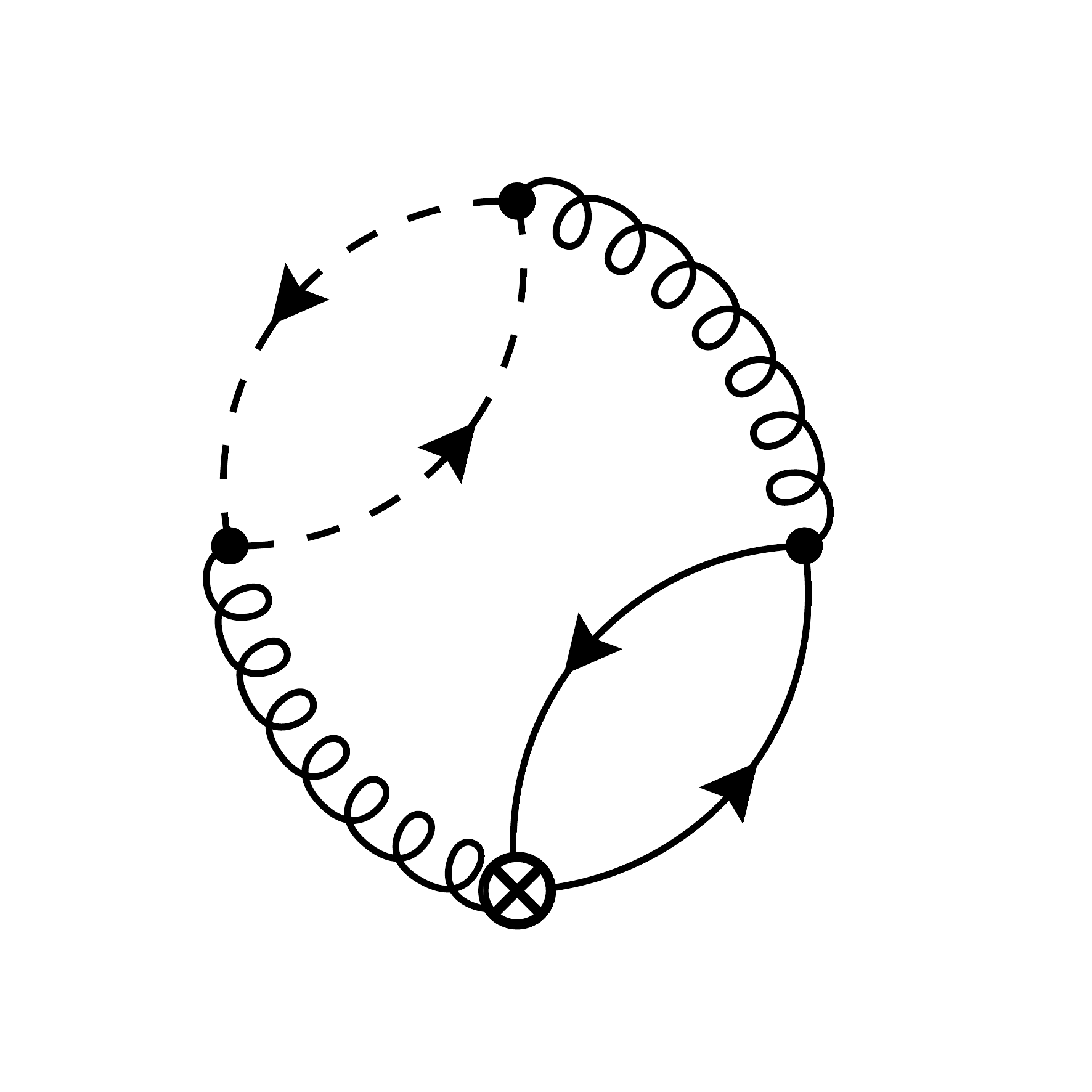} &
    \hspace*{.2em}
    \includegraphics[%
    clip,width=.3\textwidth]{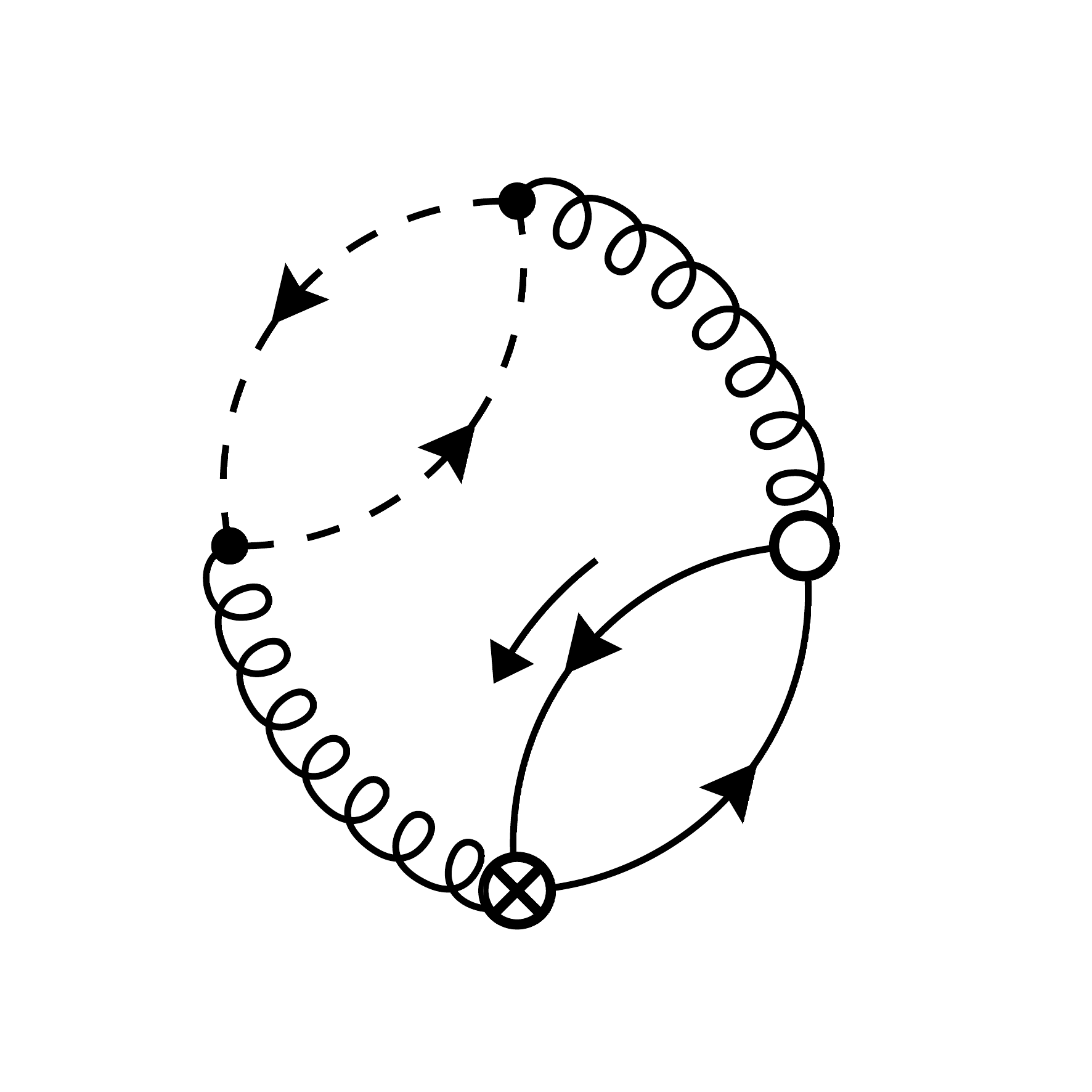}\\
    (a) &
      (b)
        \end{tabular} \parbox{.9\textwidth}{
      \caption[]{\label{fig:psidias}\sloppy
    Contributions to $\langle\bar{\chi}^\dagger(t)\slashed{D}\chi(t)\rangle$
    involving squarks.  The notation is the same as in \cref{fig:GGdias}, only
    that the vertex marked $\otimes$ now denotes the operator
    $\bar{\chi}^\dagger(t)\chi(t)$. In addition, lines with an arrow next to
    them are the associated flow lines (the arrow denotes the flow direction),
    and the symbol $\circ$ marks a flow vertex.}}  \end{center}
\end{figure}

Additional diagrams beyond pure \qcd\ which contribute to the \vev\ of
\cref{eq:dawn} in \sqcd\ occur only at \three-loop level and beyond. Two
examples are shown in \cref{fig:psidias}. They affect the \msbar\ part of the
renormalization constant as well as the finite part which characterizes the
ringed scheme. Writing
\begin{equation}\label{eq:inez}
  \begin{aligned}
    \mathring{Z}_\chi &= \zeta_\chi Z_\chi\,,
  \end{aligned}
\end{equation}
where $Z_\chi$ is the \msbar\ renormalization, and adopting the notation of
\cref{eq:Zform} for $Z_\chi$, we find for the \msbar\ coefficients up to
\nnlo\ in $\apis$ and $\apil$:
\begin{equation}\label{eq:chidiv}
  \begin{aligned}
    \gamma_{\chi,10}&=\gamma_{\chi,10}^{\text{\qcd}}\,,\qquad
    \gamma_{\chi,20}=\gamma_{\chi,20}^{\text{\qcd}}+0.1771\,\ccf\ctr\,,\\
    \gamma_{\chi,ij} &= 0\qquad\text{otherwise,}
  \end{aligned}
\end{equation}
where $\gamma_{\chi,ij}^{\text{\qcd}}$ refers to the \qcd\ results obtained in
\citeres{Luscher:2013cpa,Artz:2019bpr}:
\begin{equation}\label{eq:cern}
  \begin{aligned}
    \gamma_{\chi,10}^{\text{\qcd}} &= -\frac{3}{4}\ccf\,,\\
    \gamma_{\chi,20}^{\text{\qcd}}
    &= \left(\frac{1}{2}\ln 2-\frac{223}{96}\right)\cca\ccf
    +\left(\frac{3}{32}+\frac{1}{2}\ln 2\right)\ccf^2
    +\frac{11}{24}\ccf\ctr\nf\,.
  \end{aligned}
\end{equation}
The conversion factor to the ringed scheme reads, in our notation,
\begin{equation}\label{eq:qcdfin}
  \begin{aligned}
    \zeta_\chi = 1&
    -\apis\left(\gamma_{\chi,10}\,\lmut
    +\frac{3}{4}\ccf\ln 3+\ccf\ln 2\right)\\
    &+\apis^2\Bigg\{\frac{\gamma_{\chi,10}}{2}\left(\gamma_{\chi,10}-\betas_{20}
    \right)\lmut^2
    +\Big[\gamma_{\chi,10}\left(\betas_{20}
      -\gamma_{\chi,10}\right)\,\ln3\\
      &\quad +\frac{4}{3}\gamma_{\chi,10}\left(\betas_{20}
      -\gamma_{\chi,10}\right)\ln 2-\gamma_{\chi,20}\Big]\lmut
    +\frac{c_{\chi}^{(2)}}{16}\Bigg\}+\higher\,,
  \end{aligned}
\end{equation}
where $\gamma_{\chi,ij}$ and $\betas_{ij}$ are given in
\cref{eq:cern,eq:ceyx}.

For the non-logarithmic term at \nnlo, we find
\begin{equation}\label{eq:febe}
  \begin{aligned}
      c_\chi^{(2)} & =c_{\chi,\text{\qcd}}^{(2)}-0.3748\,\ccf\ctr\,,
  \end{aligned}
\end{equation}
where the \qcd\ part was obtained in \citere{Artz:2019bpr}:
\begin{equation}\label{eq:cc2}
  \begin{aligned}
    c_{\chi,\text{\qcd}}^{(2)} &= \cca\ccf\, c_{\chi,\text{A}}
    +\ccf^2\, c_{\chi,\text{F}}+\ccf\ctr\nf\, c_{\chi,\text{R}}\,,
  \end{aligned}
\end{equation}
with\footnote{The factor $-1/18$ should read $1/18$ in Eq.\,(B.3) of
\citere{Artz:2019bpr} (Eq.\,(131) in the \texttt{arXiv} version).}
\begin{equation}\label{eq:cc2num}
  \begin{aligned}
    c_{\chi,\text{A}} &= -23.7947,\qquad c_{\chi,\text{F}}= 30.3914,\qquad\\
    c_{\chi,\text{R}} &=
    -\frac{131}{18}+\frac{46}{3}\zeta_2
    +\frac{944}{9}\ln 2+\frac{160}{3}\ln^2 2
    -\frac{172}{3}\ln 3+\frac{104}{3}\ln2 \ln 3\\
    &-\frac{178}{3}\ln^2 3+\frac{8}{3}\text{Li}_2(1/9)
    -\frac{400}{3}\text{Li}_2(1/3)+\frac{112}{3}\text{Li}_2(3/4)
    =-3.92255\ldots\,,
  \end{aligned}
\end{equation}
where $\mathrm{Li}_2(x)$ is the dilogarithm.

\subsection{The flowed scalar field renormalization}\label{sec:scalar}

\begin{figure}
  \begin{center}
    \begin{tabular}{cccc}
      \includegraphics[%
            clip,width=.21\textwidth]%
                          {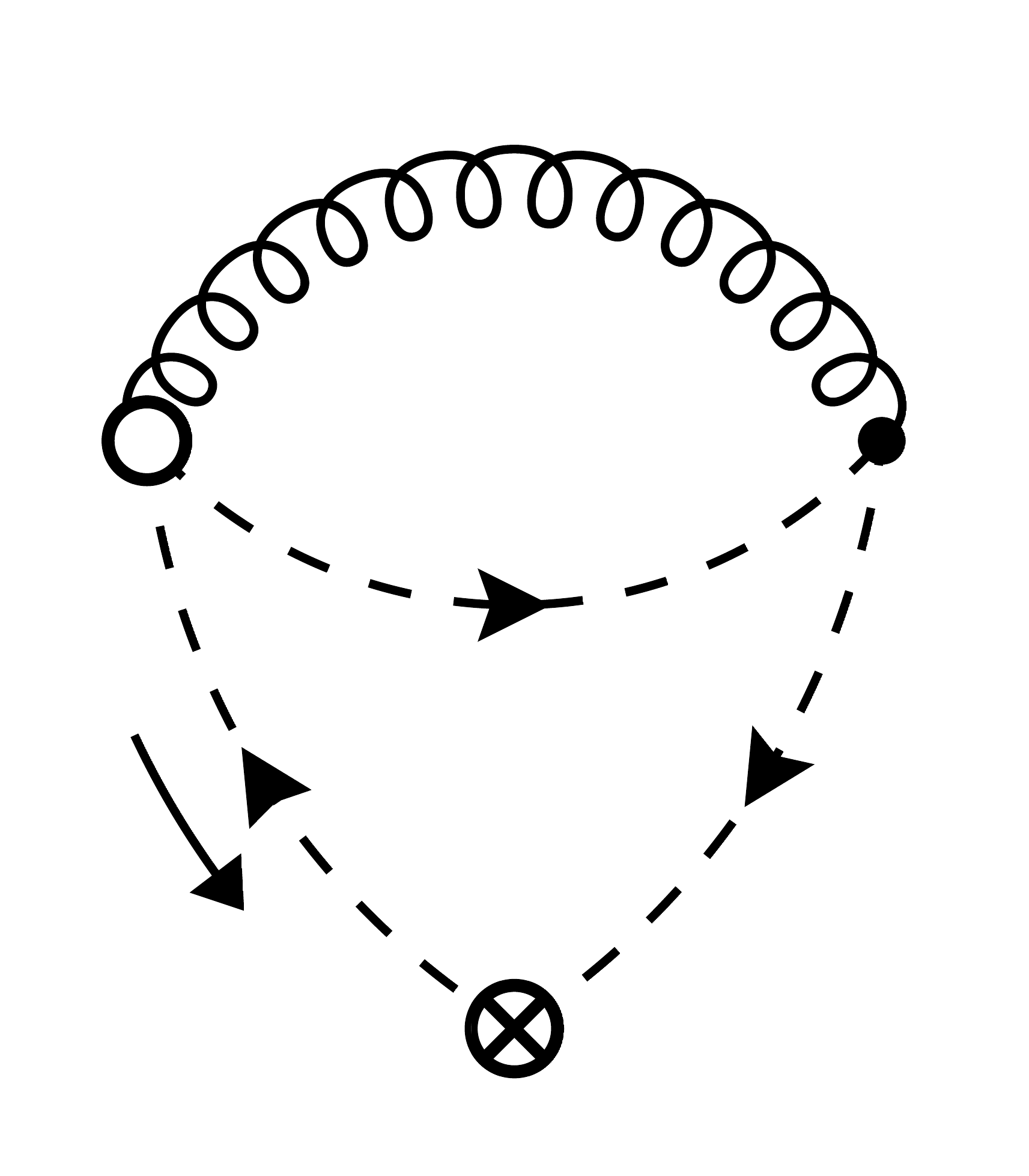}
                          &
          \includegraphics[%
            clip,width=.21\textwidth]%
                          {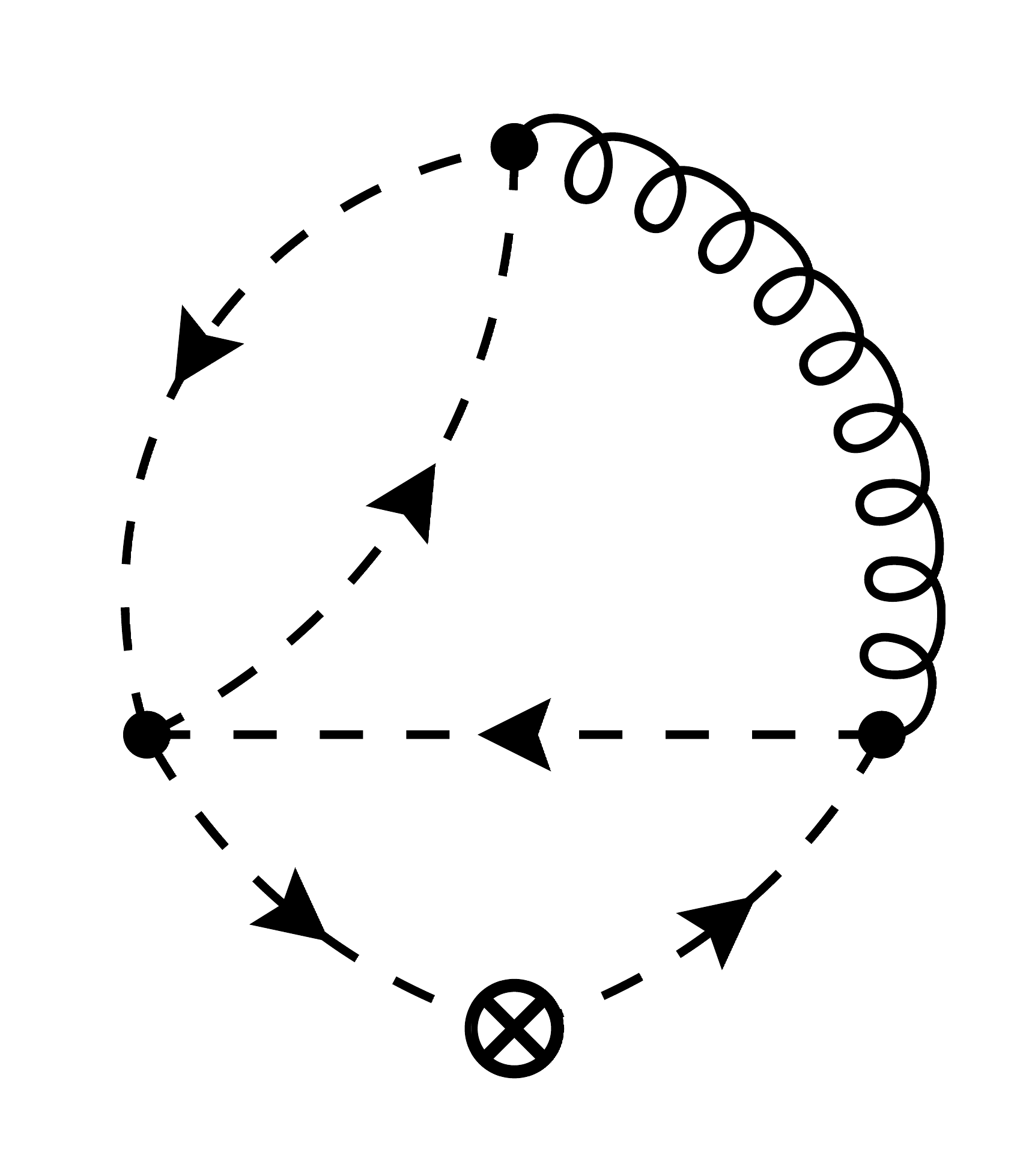}
                          &
          \includegraphics[%
            clip,width=.21\textwidth]%
                          {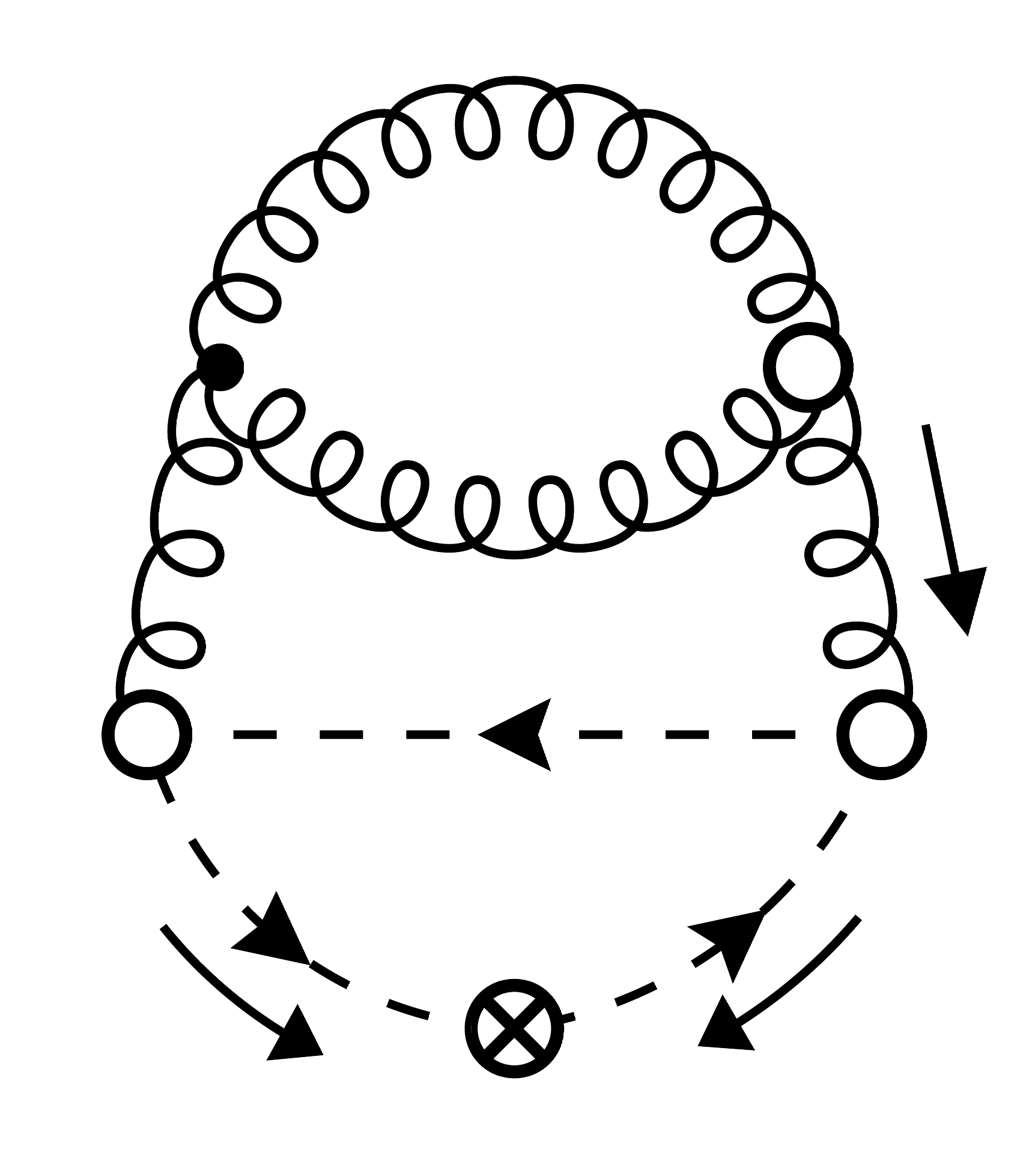}
                          &
          \includegraphics[%
            clip,width=.21\textwidth]%
                          {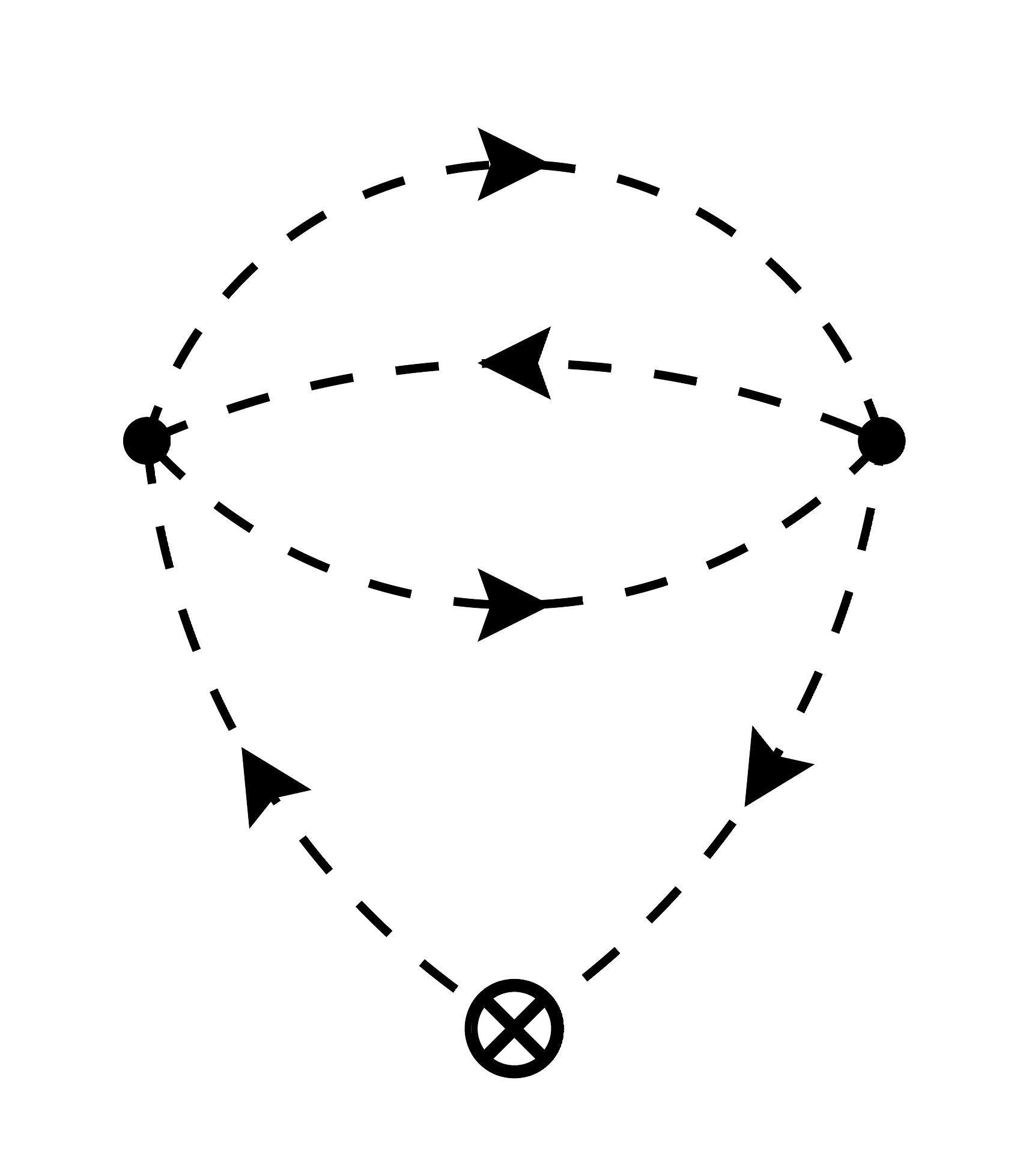}\\
                           (a) & (b) & (c) & (d)
    \end{tabular}
    \parbox{.9\textwidth}{
      \caption[]{\label{fig:phidias}\sloppy Contributions to
        $\langle\varphi^\dagger(t)\varphi(t)\rangle$ at order $\apis$,
        $\apis\apil$, $\apis^2$, and $\apil^2$ (a--d).  The notation is the
        same as in \cref{fig:GGdias}.  Lines with an arrow next to them denote
        flow lines. All ``mixed'' diagrams with both a quartic scalar and a
        gauge vertex such as diagram~(b) evaluate to zero
        individually. Diagram~(d) is the only non-vanishing \three-loop
        diagram without a gauge coupling.}}
  \end{center}
\end{figure}

Adopting the renormalization scheme defined in \cref{eq:hark} for the scalar
quark, we need to compute the scalar squark density up to \three-loop level
when aiming for the \nnlo\ result. The relevant Feynman rules for the operator
$\varphi^\dagger(t)\varphi(t)$ is the same as for the regular operator
$\phi^\dagger\phi$:
\begin{equation}\label{eq:diol}
  \begin{aligned}
    \raisebox{-5em}{\includegraphics[%
        width=.25\textwidth]{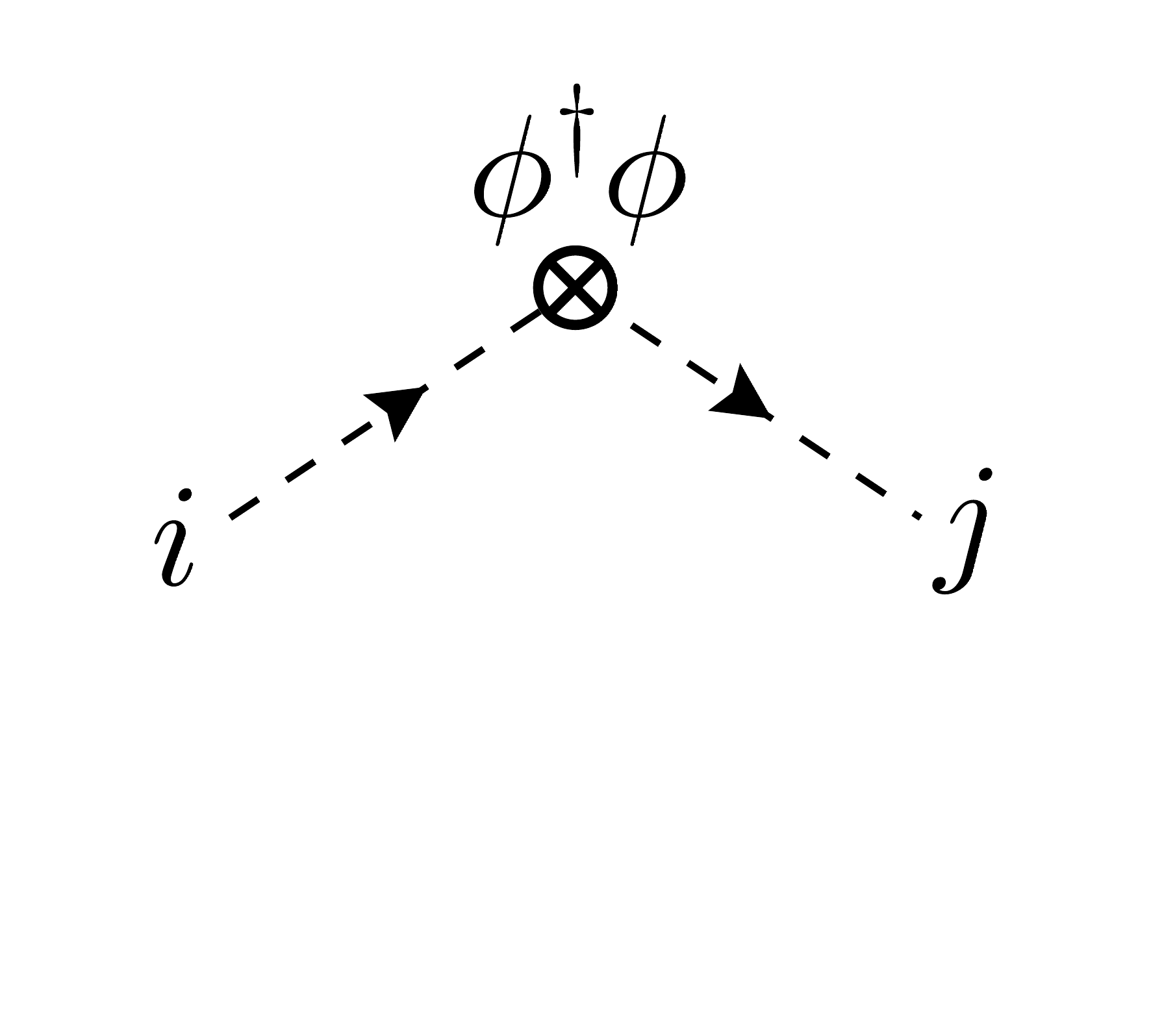}}
    & =\delta_{ij}\,.
  \end{aligned}
\end{equation}
Sample diagrams contributing at \nlo\ and \nnlo\ are shown in
\cref{fig:phidias}. In analogy to the fermionic case, we factorize the ringed
renormalization constant into an \msbar\ part $Z_\varphi$, and a conversion
factor $\zeta_\varphi$:
\begin{equation}\label{eq:craw}
  \begin{aligned}
    \mathring{Z}_\varphi &= \zeta_\varphi Z_\varphi\,.
  \end{aligned}
\end{equation}
Adopting again \cref{eq:Zform}, we find for the \msbar\ coefficients through
\nnlo:
\begin{equation}\label{eq:idle}
  \begin{aligned}
    \gamma_{\varphi,20} &= \ccf^2\Bigl(-\frac{3}{32}+\frac{11}{4}\ln2-\frac{9}{8}\ln3\Bigr)+\ccf\cca\Bigl(-\frac{65}{64}-\frac{7}{4}\ln2+\frac{9}{8}\ln3\Bigr)\\
    &+\ccf\ctr\Bigl(\frac{1}{32}+\frac{1}{8}\nf\Bigr)\,,\\
    \gamma_{\varphi,02} &= -\frac{1}{128}\left(\nc+1\right)\,,\qquad
    \gamma_{\varphi,ij} = 0\quad\text{otherwise.}
  \end{aligned}
\end{equation}
This means that the squark field renormalization is only required at the
\two-loop level and beyond. The color factor $(\nc+1)$ of \cref{eq:idle}
arises from the contraction of the \four-squark vertex with another pair of
Kronecker-$\delta$s; an overall factor $\nc$ is taken into account in
\cref{eq:hark}. Out of 377 \three-loop diagrams, only the one shown in
\cref{fig:phidias}\,(b) contributes to $\gamma_{\varphi,02}$. Remarkably,
diagrams with both a scalar and a gauge coupling vanish individually at
\three-loop level, so that $\gamma_{\varphi,11}=0$ as indicated in
\cref{eq:idle}.

For the conversion factor to the ringed scheme, we find
\begin{equation}\label{eq:coos}
  \begin{aligned}
    \zeta_\varphi &=
    1-\apis\ccf\left(1+2\ln2\right)
    +\apis^2\left\{c_{\varphi,20}^{(2)}-\lmut\left[\gamma_{\varphi,20} + \betas_{20}\ccf(1+2\ln 2)\right]\right\}\\
    &\hspace{1em}+\apil^2\,\big[c_{\varphi,02}^{(2)}
      - \lmut\gamma_{\varphi,02}\big]+ \higher\,,
  \end{aligned}
\end{equation}
where $\betas_{20}$ is given in \cref{eq:ceyx}, and
\begin{equation}\label{eq:anum}
  \begin{aligned}
    c_{\varphi,20}^{(2)}&=3.2133\ccf^2-4.268\cca\ccf
      +\ccf\ctr\left(1.086\nf+0.4547\right)\,,\\
    c_{\varphi,02}^{(2)}&=0.025391\left(\nc+1\right)\,.
  \end{aligned}
\end{equation}
The logarithmic term is consistent with the \rge
\begin{equation}\label{eq:afar}
  \begin{aligned}
    \mu^2\dderiv{}{}{\mu^2}\mathring{Z}_\varphi
    \langle\varphi^\dagger(t)\varphi(t)\rangle = 0\,,
  \end{aligned}
\end{equation}
from which one derives
\begin{equation}\label{eq:conc}
  \begin{aligned}
    \left[\deriv{}{}{\lmut}
      + \betas(\alpha_s,\apil)\deriv{}{}{\alpha_s}
      + \betal(\alpha_s,\apil)\deriv{}{}{\apil}
      - \gamma_\varphi(\alpha_s,\apil)\right]\zeta_\varphi = 0\,,
  \end{aligned}
\end{equation}
where $\gamma_\varphi$ is defined according to the generic form
of \cref{eq:gammaform}, with the coefficients given in \cref{eq:idle}.

\section{Short-flow-time expansions}\label{sec:sftx}

As an application of our results, we will derive the \sftx\ of two
bilinear flowed squark operators.

\subsection{The scalar bilinear}\label{sec:sftx_scalar}

\begin{figure}
  \begin{center}
    \begin{tabular}{cccc}
      \raisebox{0em}{%
          \includegraphics[%
            clip,width=.21\textwidth]%
                          {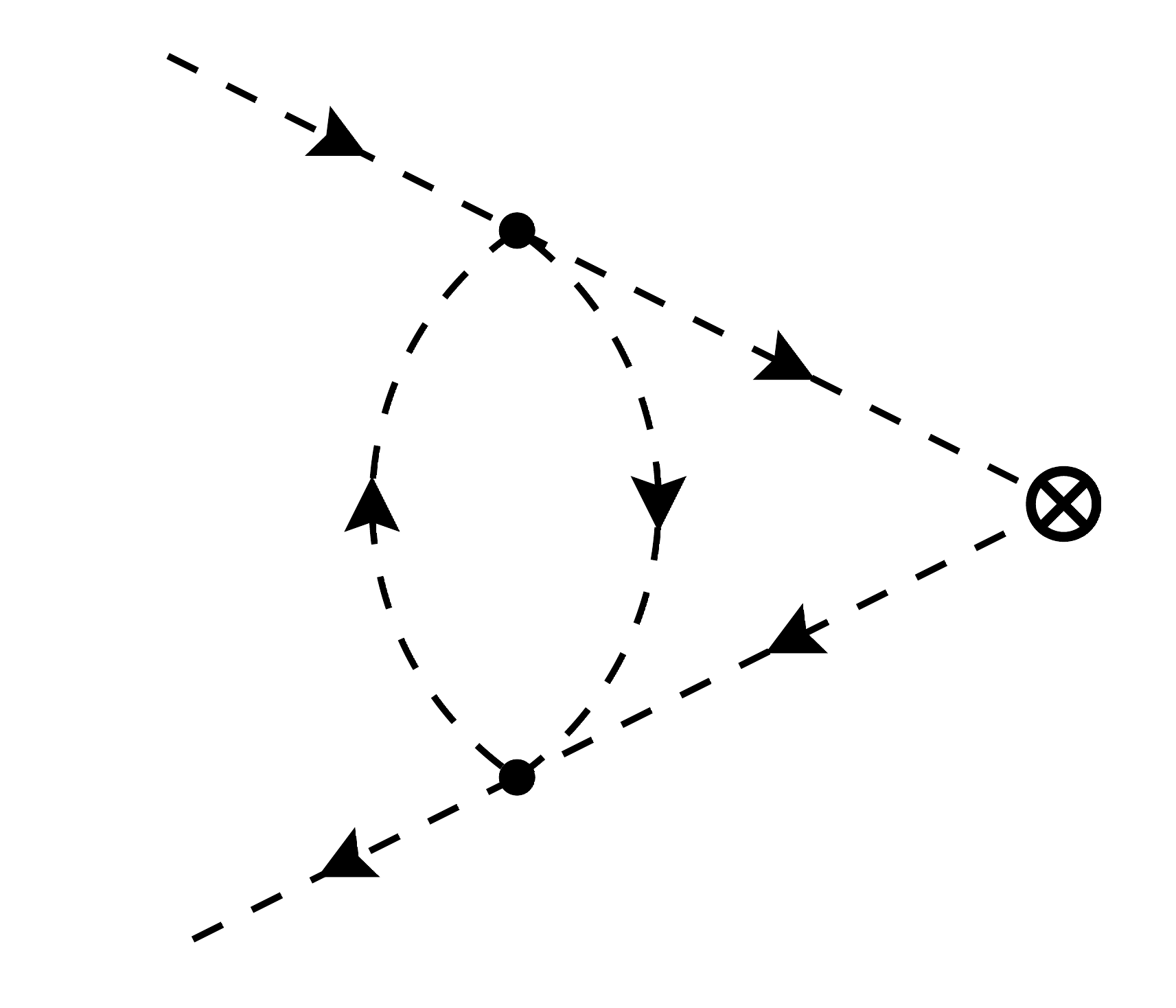}}
      &
      \raisebox{0em}{%
          \includegraphics[%
            clip,width=.21\textwidth]%
                          {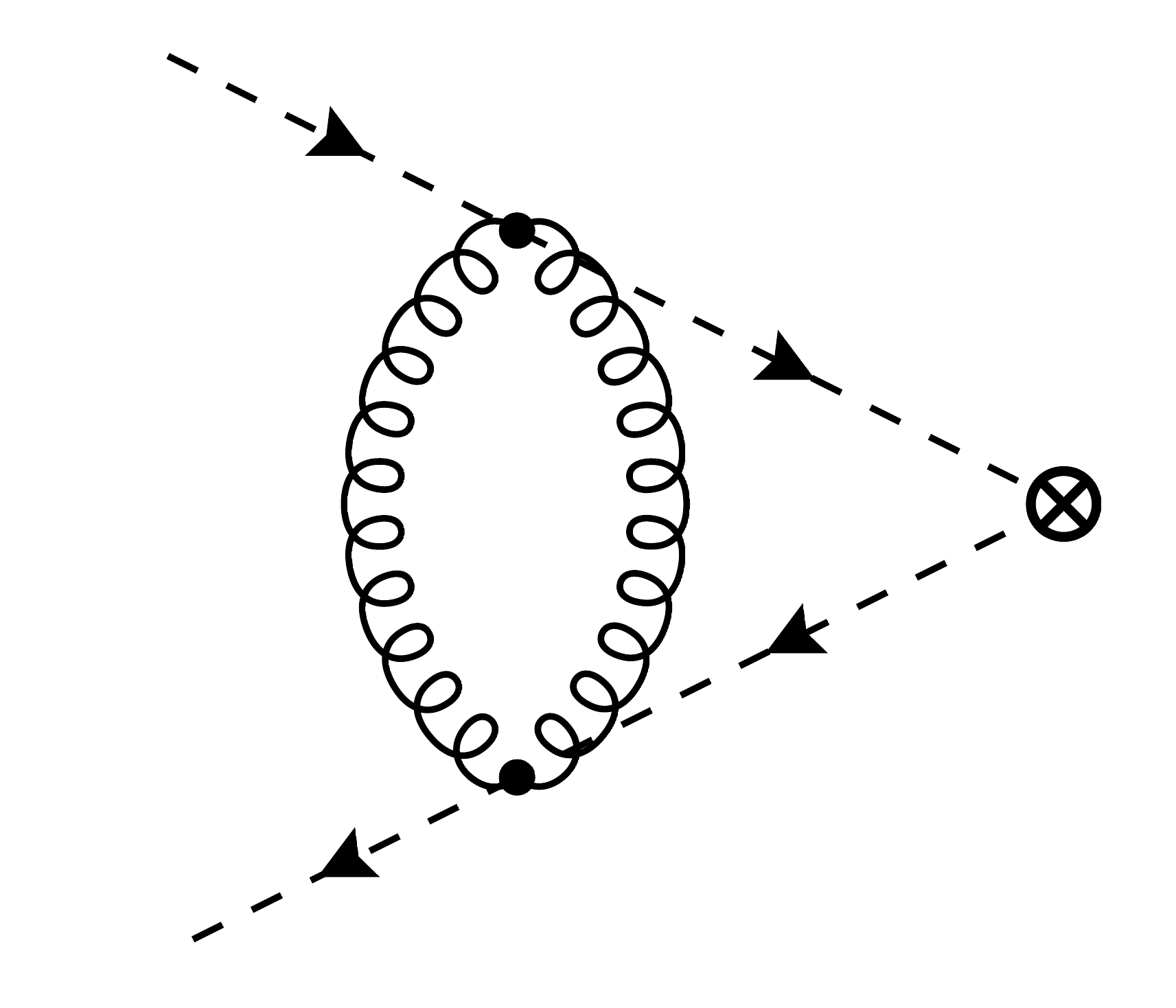}}
      &
      \raisebox{0em}{%
          \includegraphics[%
            clip,width=.21\textwidth]%
                          {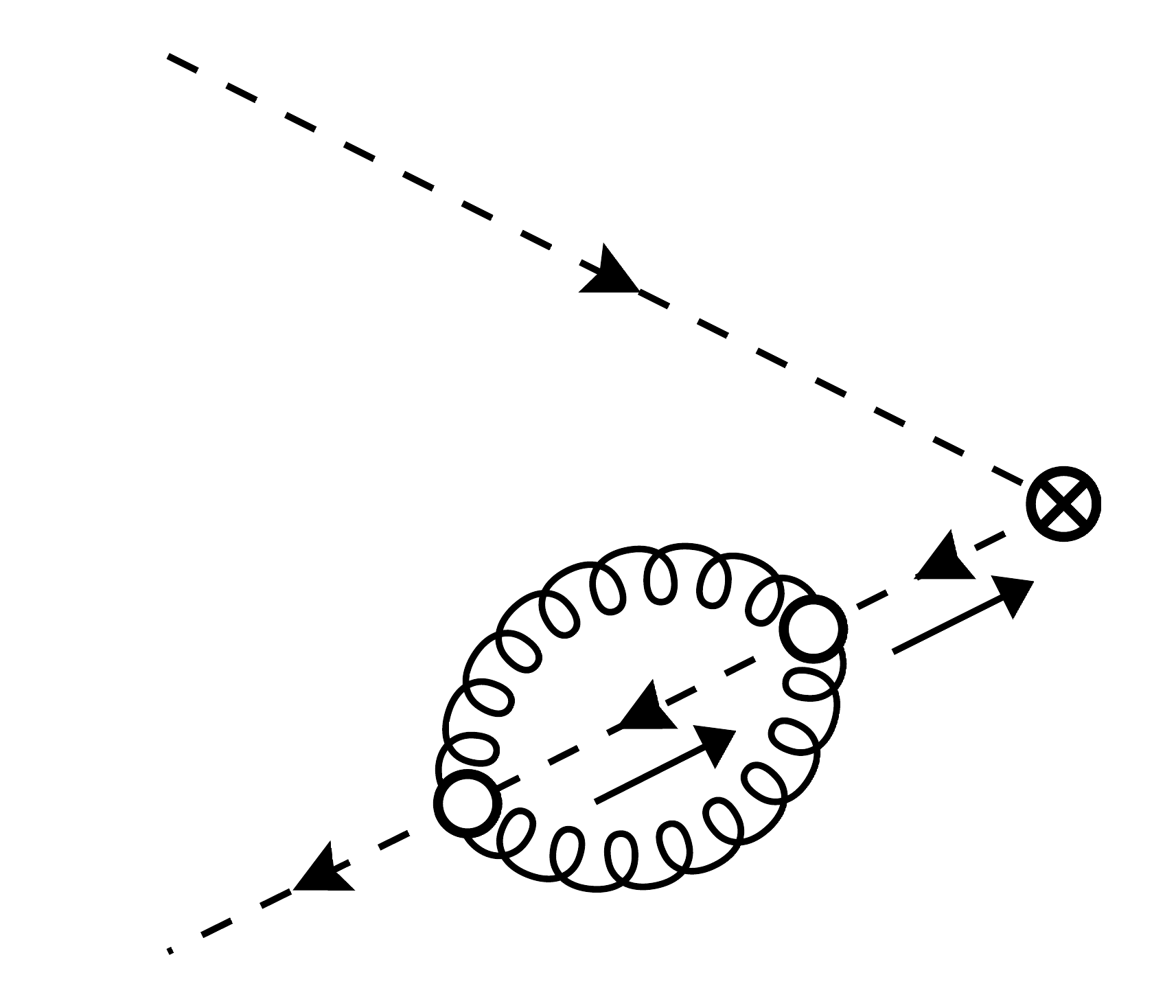}}
      &
      \raisebox{0em}{%
          \includegraphics[%
            clip,width=.21\textwidth]%
                          {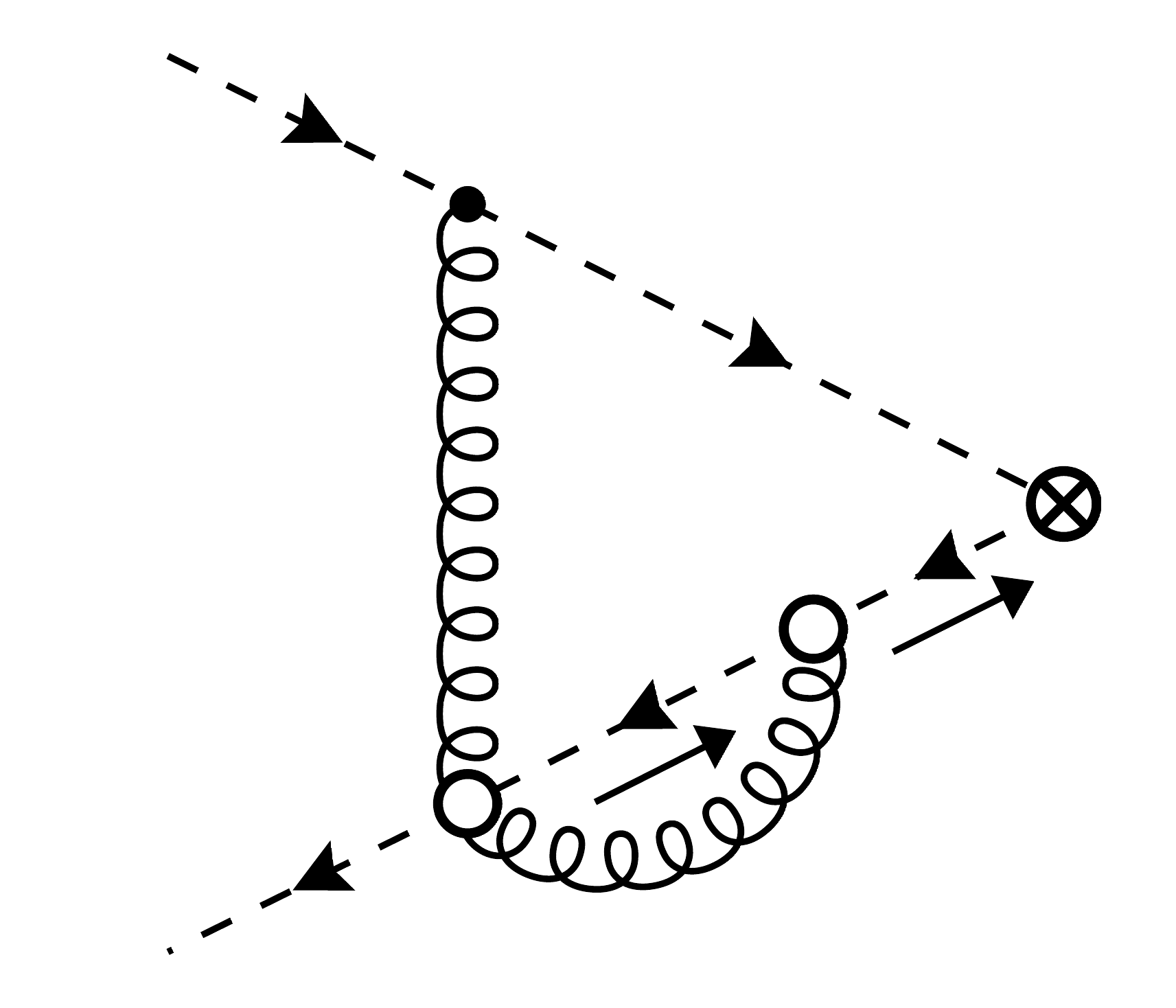}}\\
      (a) & (b) & (c) & (c)
    \end{tabular}
    \parbox{.9\textwidth}{
      \caption[]{\label{fig:tOppxOpp}\sloppy Sample diagrams contributing to
        the \sftx\ expansion of $\varphi^\dagger(t)\varphi(t)$ at \two-loop
        level. In total, we find 532~diagrams at this order (11 at \one-loop
        level), including those which vanish due to closed flow-time loops.
    }}
  \end{center}
\end{figure}

The first operator we consider is the scalar squark bilinear. Neglecting
linear and higher terms in $t$, its \sftx\ takes the form
\begin{equation}\label{eq:haag}
  \begin{aligned}
    \mathring{Z}_\varphi\varphi^\dagger(t)\varphi(t)
    = \frac{1}{t}c_{0}(t)\mathds{1} +
    c_{1}(t)Z_{\phi^\dagger\phi}\phi^\dagger\phi\,.
  \end{aligned}
\end{equation}
The coefficients $c_0(t)$ and $c_1(t)$ can be evaluated by applying the method
of projectors~\cite{Gorishnii:1983su} (see also \citere{Borgulat:2023xml}, for
example). It implies that $c_0(t)$ is obtained from the \vev\ of the operator
on the \lhs{} Using \cref{eq:hark}, it immediately follows that
\begin{equation}\label{eq:hand}
  \begin{aligned}
    c_0(t) = \frac{\nc}{32\pi^2}
  \end{aligned}
\end{equation}
to all orders in perturbation theory.  The coefficient $c_1(t)$ is determined
by the matrix element of $\varphi^\dagger\varphi$ with two external (regular)
$\phi$ fields, evaluated at zero momentum. Sample diagrams are shown in
\cref{fig:tOppxOpp}. Requiring $c_1(t)$ to be finite determines the
\msbar\ renormalization constant $Z_{\phi^\dagger\phi}$ for the unflowed
operator $\phi^\dagger\phi$ through \two-loop level. As usual, we express it
in terms of the perturbative coefficients of the anomalous dimension
$\gamma_{\phi^\dagger\phi}$, see \cref{eq:gammaform,eq:Zform}, and
find\footnote{There actually is a result for $\gamma_{\phi^\dagger\phi}$ in
the literature from more than 35 years ago\,\cite{Gorishnii:1987ik}, which,
however, disagrees with our findings.}
\begin{equation}\label{eq:bund}
  \begin{aligned}
      \gamma_{\phi^\dagger\phi,00} &= 0\,,\quad
      \gamma_{\phi^\dagger\phi,10} = \frac{3}{4}\ccf\,,\quad
    \gamma_{\phi^\dagger\phi,01} = -\frac{1}{8}\left(\nc+1\right)\,,\\
    \gamma_{\phi^\dagger\phi,20} &=
    -\frac{5}{32}\ccf^2+\frac{143}{192}\cca\ccf
    -\ccf\ctr\Bigl(\frac{71}{96}+\frac{5}{24}\nf\Bigr)\,,\\
    \gamma_{\phi^\dagger\phi,02} &= \frac{5}{128}\left(\nc+1\right)\,,\quad
    \gamma_{\phi^\dagger\phi,11} = -\frac{1}{4}\ccf\left(\nc+1\right)\,.
  \end{aligned}
\end{equation}
Since $\gamma_{\phi^\dagger\phi}$ is related to the Higgs mass anomalous dimension in the Standard Model \cite{Machacek:1984zw,Luo:2002ey}, we checked our
result by reproducing the $\text{SU}(2)$ part of the latter. 

The renormalized expression for $c_1(t)$ is then given by
\begin{equation}\label{eq:sftx:jiao}
  \begin{aligned}
  c_1(t) &=\sum_{n,m=0}^\infty c_{1,nm}\apis^n\apil^m\,,
  \end{aligned}
\end{equation}
with with coefficients up to \nnlo\ ($n+m=2$) as
\begin{equation}\label{eq:sftx:bund}
  \begin{aligned}
    c_{1,00} &= 1\,,\quad
      c_{1,10} = -\ccf\left(\frac{5}{4}+2\ln2\right)\,,\quad
      c_{1,01} = \frac{1}{8}\left(\nc+1\right)\,,\\
      c_{1,20} &=
      \ccf^2\biggl(-\frac{1}{32}-\frac{5}{32}\zeta_2+\frac{47}{8}\ln2
      -\frac{7}{8}\ln^22+\frac{9}{8}\ln2\ln3-\frac{39}{16}\ln3
      -\frac{9}{32}\ln^23\\&
      +\frac{15}{8}\text{Li}_2\left(1/4\right)\biggr)
      +\cca\ccf\biggl(-\frac{397}{192}-\frac{7}{64}\zeta_2-\frac{47}{8}\ln2
      +\frac{11}{8}\ln^22-\frac{9}{8}\ln2\ln3\\&
      +\frac{75}{16}\ln3+\frac{9}{32}\ln^23
      +\frac{3}{4}\text{Li}_2\left(1/4\right)\biggr)
      +\ccf\ctr\biggl(-\frac{55}{96}-\frac{1}{32}\zeta_2
      +\frac{11}{3}\ln2\\&-\frac{3}{2}\ln3
      +\frac{7}{8}\text{Li}_2\left(1/4\right)\biggr)
      +\ccf\ctr\nf\biggl(\frac{5}{12}+\frac{1}{8}\zeta_2\biggr)
      +c_{\varphi,20}^{(2)}\,,\\
      c_{1,02} &= (\nc+1)\biggl(-\frac{15}{128}-\frac{3}{128}\zeta_2
      +c_{\varphi,02}^{(2)}\biggr)\,,\\
      c_{1,11} &= \ccf\left(\nc+1\right)\biggl(\frac{3}{8}
      +\frac{3}{8}\ln2-\frac{9}{16}\ln3
      -\frac{3}{8}\text{Li}_2\left(1/4\right)\biggr)\,,
  \end{aligned}
\end{equation}
where $c_{\varphi,02}^{(2)}$ and $c^{(2)}_{\varphi,20}$ are given
in \cref{eq:anum}, and we have set $\mu=\mu_t$ for the sake of brevity,
see \cref{eq:abed}. The result for general values of $\mu$ can be
reconstructed from the relation
\begin{equation}\label{eq:sftx:gogo}
  \begin{aligned}
    \mu^2\dderiv{}{}{\mu^2}c_1(t) &=
    \left[\deriv{}{}{\lmut} +
      \betas(\apis,\apil)\deriv{}{}{\apis}
      + \betal(\apis,\apil)\deriv{}{}{\apil}\right]c_1(t) =
    \gamma_{\phi^\dagger\phi}\,c_1(t)\,,
  \end{aligned}
\end{equation}
which follows from the renormalization scale independence of the \lhs\ of
\cref{eq:haag}. For convenience of the reader,
we include the explicitly $\mu$-dependent terms in the ancillary file
accompanying this paper, see \cref{sec:ancillary}.

\subsection{The Noether current}\label{sec:sftx_noether}

The second example we consider is the Noether current for the squark, given by
the operator $\phi^\dagger \overleftrightarrow{D}_\mu\phi$. Its Feynman rules
are
\begin{equation}\label{eq:diolx}
  \begin{aligned}
    \raisebox{-5em}{\includegraphics[%
        width=.3\textwidth]{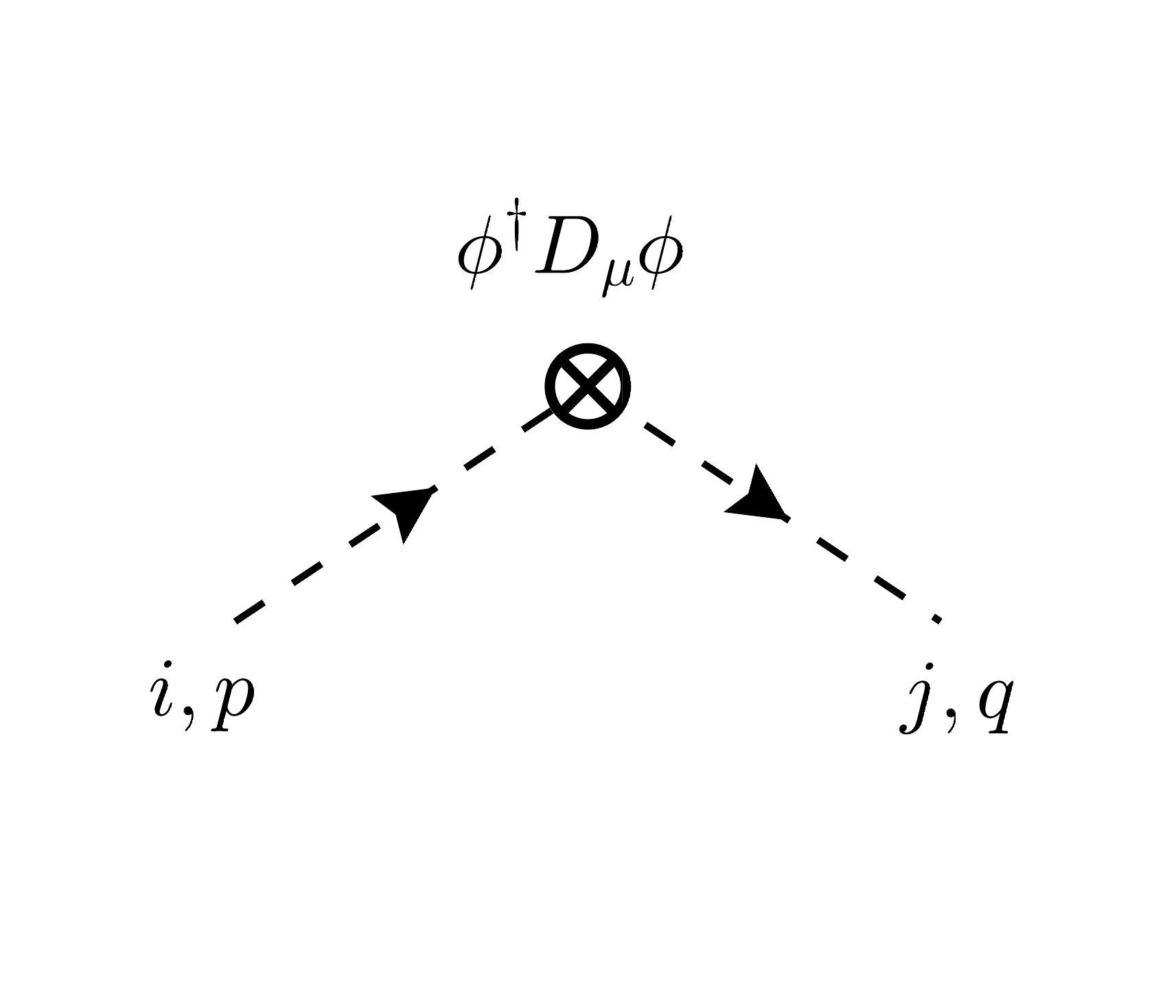}}
    & =i\delta_{ij}(p-q)_\mu\,,\\[-1em]
    \raisebox{-5em}{\includegraphics[%
        width=.3\textwidth]{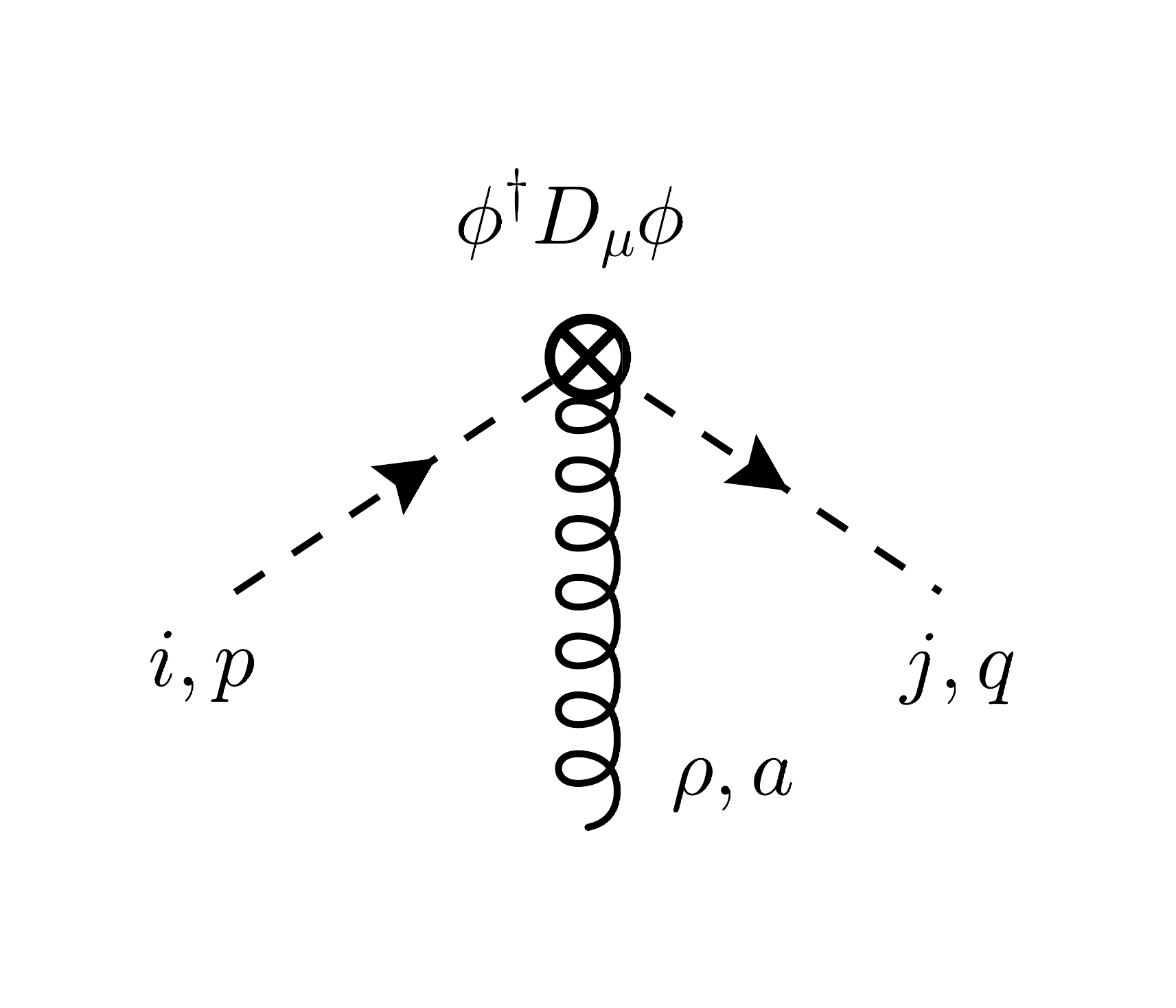}}
    & =2\,t^a_{ji} \delta_{\mu\rho}\,,
  \end{aligned}
\end{equation}
where the momenta are assumed outgoing.

Neglecting again higher orders in $t$, the \sftx\ of the flowed version of
this operator reads
\begin{equation}\label{eq:sftn}
  \begin{aligned}
    \mathring{Z}_\varphi\varphi^\dagger(t)\overleftrightarrow{D}_\mu\varphi(t) =
    d(t)\phi^\dagger\overleftrightarrow{D}_\mu\phi\,,
  \end{aligned}
\end{equation}
with the coefficients up to \nnlo\ given by
\begin{equation}\label{eq:dt}
  \begin{aligned}
    d(t) &=\sum_{n,m=0}^\infty d_{nm}\apis^n\apil^m\,,\\
    d_{00} &= 1\,,\quad
    d_{10} = -\ccf\left(1+2\ln2\right)\,,\quad d_{01} = 0\,,\\
    d_{20} &=
    \ccf^2\biggl(\frac{111}{128}-\frac{1}{8}\zeta_2
    +\frac{13}{2}\ln2-\frac{7}{8}\ln^22+\frac{9}{8}\ln2\ln3
    -\frac{33}{8}\ln3-\frac{9}{32}\ln^23\\&
    +\frac{3}{2}\text{Li}_2\left(1/4\right)\biggr)
    +\cca\ccf\biggl(-\frac{231}{256}-\frac{1}{8}\zeta_2
    -\frac{19}{8}\ln2+\frac{11}{8}\ln^22-\frac{9}{8}\ln2\ln3\\&
    +\frac{21}{16}\ln3+\frac{9}{32}\ln^23
    +\frac{9}{32}\text{Li}_2\left(1/4\right)\biggr)
    +\ccf\ctr\biggl(\frac{7}{128}+\frac{5}{32}\nf\biggr)
    +c_{\varphi,20}^{(2)}\\
    & + d_{10}\betas_{20}\lmut\,,\\
    d_{02} &= (\nc+1)\biggl(-\frac{7}{512}+c_{\varphi,02}^{(2)}\biggr)\,,\\
    d_{11} &= \ccf\biggl(-\frac{3}{16}-\frac{1}{32}\zeta_2
    +\frac{9}{32}\text{Li}_2\left(1/4\right)\biggr)\,,
  \end{aligned}
\end{equation}
where $c_{\varphi,20}^{(2)}$, $c_{\varphi,02}^{(2)}$, and $\betas_{20}$ are
given in \cref{eq:anum,eq:ceyx}, respectively.  The logarithmic term is
consistent with
\begin{equation}\label{eq:sftx:gabo}
  \begin{aligned}
    \mu^2\dderiv{}{}{\mu^2}d(t) = 0\,,
  \end{aligned}
\end{equation}
which follows from the renormalization scale invariance of both the \lhs\
of \cref{eq:sftn} and the regular Noether current.  It is worth noting that
the \nlo\ coefficient $d_{10}$ arises completely from the ringed-scheme
conversion of \cref{eq:coos}, while it vanishes in the $\msbar$-scheme.

\section{Flowed anomalous dimension}\label{sec:fanom}

The gradient flow can be viewed as a renormalization scheme. The corresponding
renormalization group equation for a set of flowed operators
$\tilde{\mathcal{O}}=(\tilde{\mathcal{O}}_1,\cdots)$ in this scheme can be
written as
\begin{equation}\label{eq:fano:kino}
  \begin{aligned}
    t\dderiv{}{}{t}\tilde{\mathcal{O}}(t) &=
    \tilde{\gamma}\tilde{\mathcal{O}}(t)\,,
  \end{aligned}
\end{equation}
where the flowed anomalous dimension matrix $\tilde{\gamma}$ can be expressed
in terms of the matching matrix $\zeta(t)$ of the \sftx, defined as
\begin{equation}\label{eq:fano:fret}
  \begin{aligned}
    \tilde{\mathcal{O}}(t) &= \zeta(t)\,\mathcal{O}\,.
  \end{aligned}
\end{equation}
Setting $\mu^2\sim 1/t$, one
finds~\cite{Harlander:2020duo,Borgulat:2023xml} 
\begin{equation}\label{eq:fano:abby}
  \begin{aligned}
    \tilde{\gamma} &= \left[\left(t\deriv{}{}{t}
      - \betas\deriv{}{}{\apis} - \betal\deriv{}{}{\apil}\right)\zeta(t)\right]
    \zeta^{-1}(t) +\zeta(t)\gamma\zeta^{-1}(t)\,,
  \end{aligned}
\end{equation}
were the partial derivative \wrt\ $t$ takes into account any power-dependence
of $\zeta(t)$ on $t$, and $\gamma$ is the anomalous dimension of the regular
operator $\mathcal{O}$, i.e.\footnote{Note that \citere{Borgulat:2023xml}
defines $\gamma$ with the opposite sign.}
\begin{equation}\label{eq:fano:kane}
  \begin{aligned}
    \mu^2\dderiv{}{}{\mu^2}\mathcal{O} &= -\gamma\mathcal{O}\,.
  \end{aligned}
\end{equation}
For the Noether current in \cref{eq:sftn}, we have
\begin{equation}\label{eq:fano:cone}
  \begin{aligned}
    \tilde{\mathcal{O}}(t) &=
    \mathring{Z}_\varphi\varphi^\dagger(t)\overleftrightarrow{D}
    \varphi(t)\,,\qquad&
    \zeta(t) &= d(t)\,,\qquad \gamma=0\,,\\
    \mathcal{O}(t) &=
    \phi^\dagger\overleftrightarrow{D}\phi(t)\,,&
    t\dderiv{}{}{t}\tilde{\mathcal{O}}(t) &=
    \tilde{\gamma}_d\tilde{\mathcal{O}}(t)\,.
  \end{aligned}
\end{equation}
In this case, the first term in brackets of \cref{eq:fano:abby} does not
contribute. Given the result in \cref{eq:dt}, we can determine
$\tilde{\gamma}_d$ through \order{\apis^i\apil^j} with $i+j=3$:
\begin{equation}\label{eq:fano:erik}
  \begin{aligned}
    \tilde{\gamma}_d &=\sum_{n,m=0}^\infty \tilde{\gamma}_{d,nm}
    \apis^n\apil^m\,,\\
    \tilde{\gamma}_{d,00}
    &=\tilde{\gamma}_{d,10}
    =\tilde{\gamma}_{d,01}
    =\tilde{\gamma}_{d,11} = 0\,,\\
    \tilde{\gamma}_{d,20} &= \betas_{20}\,d_{10}\,,\\
    \tilde{\gamma}_{d,30} &= 
      \betas_{30}\,d_{10} + \betal_{20}\,d_{11} - \betas_{20}\,(d_{10}^2 -
      2\,d_{20})\,,
      \\
    \tilde{\gamma}_{d,21} &= 
    2\,\betal_{20}\,d_{02}
    + (\betal_{11} + \betas_{20})\,d_{11}\,,\\
    \tilde{\gamma}_{d,12} &= 
      2\,\betal_{11}\,d_{02} + \betal_{02}\,d_{11}\,,\\
    \tilde{\gamma}_{d,03} &= 
    2\,\betal_{02}\,d_{02} \,.
  \end{aligned}
\end{equation}
In the above relations, the renormalization scale is chosen to be $\mu=\mu_t$,
such that $\lmut=0$, see \cref{eq:abed}; the full relation can be found in the
ancillary file to this paper, see \cref{sec:ancillary}.

For the scalar bilinear of \cref{eq:haag}, on the other hand, we need to take
into account the mixing with the unit operator. Therefore, in this case we set
\begin{equation}\label{eq:fano:amos}
  \begin{aligned}
    \tilde{\mathcal{O}}(t) &=
    \left(
    \begin{matrix}
      \mathds{1}\\
      \mathring{Z}_\varphi\varphi^\dagger(t)\varphi(t)
    \end{matrix}
    \right)\,,&
    \zeta(t) &= \left(
    \begin{matrix}
      1 & 0\\
      \frac{1}{t}c_0 & c_1(t)
    \end{matrix}
    \right)\,,\qquad\gamma = \gamma_{\phi^\dagger\phi}\,,\\
    \mathcal{O}(t) &=
    \left(
    \begin{matrix}
      \mathds{1}\\
      Z_{\phi^\dagger\phi}\phi^\dagger\phi
    \end{matrix}
    \right)\,,&
    t\dderiv{}{}{t}\tilde{\mathcal{O}}(t) &=
    \tilde{\gamma}_c\tilde{\mathcal{O}}(t)\,.
  \end{aligned}
\end{equation}
This leads to 
\begin{equation}\label{eq:fano:dory}
  \begin{aligned}
    \tilde{\gamma}_c &=
    \left(
    \begin{matrix}
      0 & 0 \\
      -\frac{1}{t}c_0\left(1+\tilde{\gamma}_{1}\right)
      & \tilde{\gamma}_{1}
    \end{matrix}
    \right) - \zeta(t)\gamma_{\phi^\dagger\phi}\zeta^{-1}(t)\,,
  \end{aligned}
\end{equation}
with
\begin{equation}\label{eq:fano:knag}
  \begin{aligned}
    \tilde{\gamma}_{1} &=
    -\left[\left(\betas\deriv{}{}{\apis}+\betal\deriv{}{}{\apil}\right)c_1(t)
      \right]c_1^{-1}(t)\,.
  \end{aligned}
\end{equation}
This can be easily expressed in terms of the coefficients $c_{1,ij}$,
$\betas_{ij}$, and $\betal_{ij}$ from \cref{eq:sftx:bund,eq:ceyx,eq:jimp},
similar to \cref{eq:fano:erik}.

\section{Conclusions and Outlook}\label{sec:conc}

We have presented the generalization of the \gff\ formalism to a gauge theory
with scalar particles. Aside from possible phenomenologial applications, for
example the Higgs sector of the \sm, our results validate the consistency of
the formalism in this context. We show that, similar to the fermionic case,
the flowed scalar field requires renormalization, albeit only
beyond \nlo. In addition to the \msbar\ scheme, we define a quasi-physical
scheme for the scalar quarks, defined at non-zero flow-time, and analogous to
the ringed scheme for fermions. It leads to renormalization group invariant
Green's functions of the scalar field and can be implemented both
perturbatively and on the lattice. We evaluate the corresponding
renormalization constant through two loops.

Our results constitute an important step towards a gradient-flow description
of the full \sm, which paves the way for novel perspectives on calculations
within \smeft, for example.

\paragraph{Acknowledgments.}
We thank Andrei Kataev for helpful communication concerning the disagreement
of \cref{eq:bund} with \citere{Gorishnii:1987ik}.  Furthermore, we would like
to thank Jonas Rongen and Aiman el Assad for preliminary work on this project.
This research was supported by the \textit{Deutsche Forschungsgemeinschaft
(DFG, German Research Foundation)} under grant 460791904.

\begin{appendix}

\section{Renormalization constants}\label{sec:renormalization}

\begin{figure}
  \begin{center}
    \begin{tabular}{c}
          \includegraphics[%
            clip,width=.3\textwidth]%
                          {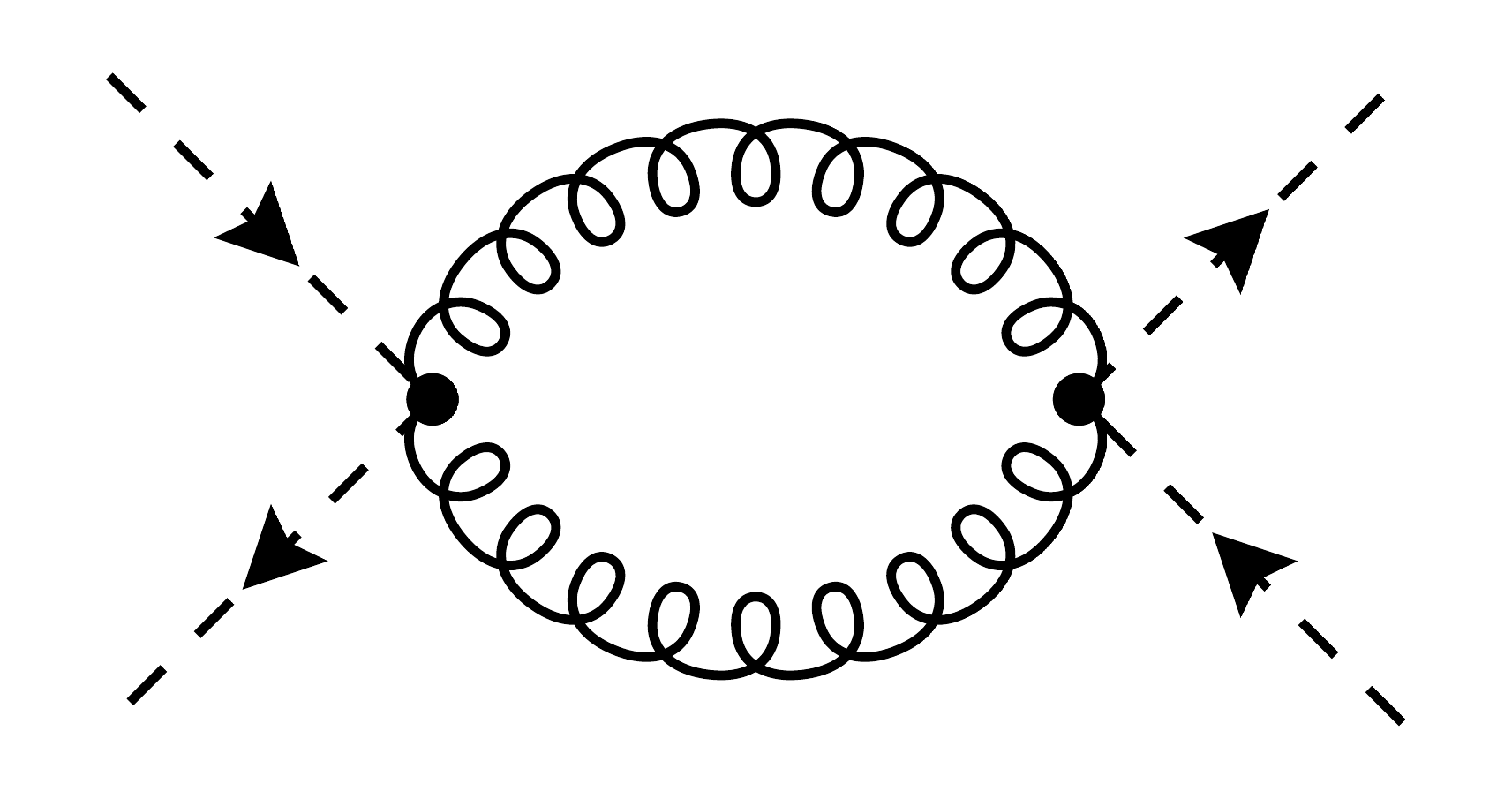}
    \end{tabular}
    \parbox{.9\textwidth}{
      \caption[]{\label{fig:Zslambda}\sloppy Example diagram for the origin of
        the non-multiplicative renormalization $Z_{\text{s}\lambda}$ of the
        scalar coupling $\apil$.  }}
  \end{center}
\end{figure}

The couplings are renormalized by
\begin{equation}\label{eq:coupdef}
  \begin{aligned}
    \apis^\bare&=\biggl(\frac{\mu^2e^{\EulerGamma}}{4\pi}\biggl)^\ep
    Z_\text{s}(\apis,\apil)\,\apis\,,\\
    \apil^\bare&=\biggl(\frac{\mu^2e^{\EulerGamma}}{4\pi}\biggl)^\ep
    \Bigl[Z_{\text{s}\lambda}(\apis)\apis+Z_{\lambda}(\apis,\apil)\apil\Bigr]\,,
  \end{aligned}
\end{equation}
where the dependence of the renormalized couplings $\apis$ and $\apil$ on the
renormalization scale $\mu$ has been suppressed. It is governed by the
renormalization group equations
\begin{equation}\label{eq:betadef}
  \begin{aligned}
    \mu^2\frac{d}{d\mu^2}\apis&=\betas(\apis,\apil)
    \equiv-\ep\,\apis
    -\sum_{n=2}^\infty\sum_{i=0}^{n} \betas_{i,n-i}
    \apis^i\,\apil^{n-i}\,,\\
    \mu^2\frac{d}{d\mu^2}\apil&=\betal(\apis,\apil)
    \equiv
    -\ep\,\apil
    -\sum_{n=2}^\infty\sum_{i=0}^{n} \betal_{i,n-i}
    \apis^i\,\apil^{n-i}\,.
  \end{aligned}
\end{equation}
The coefficients for the strong beta function through \two-loop level,
i.e.\ $n=3$, are~\cite{Jones:1975127}
\begin{equation}\label{eq:ceyx}
  \begin{aligned}
    \betas_{20} &=
    \frac{1}{4}\left(\frac{11}{3}\cca - \frac{4}{3}\nf\ctr
    -\frac{1}{3}\ctr\right)\,,\\
    \betas_{30} &= \frac{1}{16}\left[\frac{34}{3}\cca^2
    -\left(4\ccf+\frac{20}{3}\cca\right)\ctr\nf
    -\left(4\ccf+\frac{2}{3}\cca\right)\ctr\right]\,,\\
    \betas_{ij} &= 0\qquad \text{otherwise,}
  \end{aligned}
\end{equation}
where $\nf$ is the number of quark flavors, and we only take into account a
single scalar quark field, see \cref{eq:sqcd}.  For the scalar coupling, the
coefficients through \two-loop level are~\cite{Gross:1973,Gorishnii:1987ik}
\begin{equation}\label{eq:jimp}
  \begin{aligned}
    \betal_{20} &= -\frac{12\ccf^2-3\cca\ccf+12\ccf\ctr}
          {4\left(\nc+1\right)}\,,\\
    \betal_{11} &= \frac{3}{2}\ccf\,,\qquad
    \betal_{02} = -\frac{1}{8}\left(\nc+1\right)-\frac{3}{8}\,,\\
    \betal_{ij} &= 0\qquad\text{otherwise.}
  \end{aligned}
\end{equation}
The renormalization constants in \cref{eq:coupdef} are then given by
\begin{equation}
  \begin{aligned}
    Z_\text{s}&=1-\apis\frac{\betas_{20}}{\ep}
    -\apis^2\left(\frac{\left(\betas_{20}\right)^2}{\ep^2}
    +\frac{\betas_{30}}{2\ep}\right)\,,\\
    Z_{\lambda}&=1-\apis\frac{\betal_{11}}{\ep}
    -\apil\frac{\betal_{02}}{\ep}\,,\qquad
    Z_{\text{s}\lambda}=-\apis\frac{\betal_{20}}{\ep}\,.
  \end{aligned}
\end{equation}
Though well known, let us emphasize the non-multiplicativity of the $\apil$
renormalization, manifest through the the non-vanishing expression for
$Z_{\text{s}\lambda}$, and caused by diagrams like the one shown
in \cref{fig:Zslambda}.

The anomalous dimension $\gamma$ of a particular quantity is related to the
corresponding renormalization constant $Z$ by
\begin{equation}\label{eq:gammadef}
  \begin{aligned}
  \mu^2\frac{d}{d\mu^2}Z&=-\gamma\,Z\,.
  \end{aligned}
\end{equation}
We write it as
\begin{equation}\label{eq:gammaform}
  \begin{aligned}
  \gamma&=-\sum_{n,m=0}^\infty\gamma_{nm}\apis^n\apil^m\,.
  \end{aligned}
\end{equation}
The generic form of an (gauge independent)
$\msbar$-renormalization constant is then given by
\begin{equation}\label{eq:Zform}
  \begin{aligned}
    Z&=1-\apis\frac{\gamma_{10}}{\ep}
    -\apil\frac{\gamma_{01}}{\ep}
    +\apis^2\left(\frac{1}{2\ep^2}\Bigl[\gamma_{10}^2
      +\betal_{20}\gamma_{01}+\betas_{20}\gamma_{10}\Bigr]
    -\frac{\gamma_{20}}{2\ep}\right)\\
    &+\apil^2\left(\frac{1}{2\ep^2}\Bigl[\gamma_{01}^2
      +\betal_{02}\gamma_{01}\Bigr]-\frac{\gamma_{02}}{2\ep}\right)\\
    &+\apis\apil\left(\frac{1}{\ep^2}\Bigl[\gamma_{01}\gamma_{10}
      +\frac{\betal_{11}\gamma_{01}}{2}\Bigr]
    -\frac{\gamma_{11}}{2\ep}\right)+\higher
  \end{aligned}
\end{equation}

\section{Ancillary file}\label{sec:ancillary}

For the reader's convenience, we provide the results of this paper in an
ancillary file in \texttt{Mathematica} format.  A list of the results and
their corresponding expressions in the ancillary file is found in
\cref{tab::results}.

The mixing coefficients $c_1$ and $d$ are provided in both the ringed scheme
of the scalar quarks as well as in the $\msbar$ scheme.  The
parameter \verb$Xzetaphi$ allows to switch between the schemes;
setting \verb$Xzetaphi=0/1$ corresponds to the $\msbar$/ringed scheme.  The
flowed anomalous dimension $\tilde{\gamma}_d$ is only given in the ringed
scheme.

The results depend on the variables listed in \cref{tab::variables}. Results
in the ringed scheme depend on the numerically known coefficients
$c_{\varphi,02}^{(2)}$ and $c_{\varphi,20}^{(2)}$ (see \cref{eq:anum}). They
can be inserted by applying the \texttt{Mathematica} replacement rule
\texttt{ReplaceC2}.

\begin{table}
  \begin{center}
    \caption{ The expressions of the ancillary file that
      encode the main results of this paper.}
    \begin{tabular}{lll}
      expression & meaning & reference\\\hline
      \verb$Et$ &  $E(t)$ &  \cref{eq:knag}\\
      \verb$zetachi$ &  $\zeta_\chi $ &  \cref{eq:qcdfin}\\
      \verb$gammaphiphi$ &  $\gamma_{\phi^\dagger\phi}$ &  \cref{eq:bund},
      \cref{eq:gammaform}\\
      \verb$Zphiphi$ &  $\mathring{Z}_{\phi^\dagger\phi}$ &
      \cref{eq:haag}\\
      \verb$gammaphi$ &  $\gamma_\varphi$ &  \cref{eq:idle},
      \cref{eq:gammaform}\\
      \verb$Zphi$ &  $\mathring{Z}_\varphi$ &  \cref{eq:craw}\\
      \verb$zetaphi$ &  $\zeta_\varphi $ &  \cref{eq:coos}\\
      \verb$c1$ &  $c_{1}(t)$ &  \cref{eq:haag}\\
      \verb$d$ &  $d(t)$ &  \cref{eq:sftn}\\
      \verb$tildegammad$ &  $\tilde{\gamma}_d$ &
      \cref{eq:fano:erik}
    \end{tabular}
  \end{center}
  \label{tab::results}
\end{table}

\begin{table}
  \begin{center}
    \caption{Notation for the variables in the
      ancillary file.}
    \begin{tabular}{lll}
      symbol & meaning & definition\\\hline
      \verb$nc1$ &  $\nc+1$ &  \cref{sec:ren}\\
      \verb$tr$ &  $\ctr$ &   \cref{eq:fahr}\\
      \verb$cf$ &  $\ccf$ &  \cref{eq:acti:kahn}\\
      \verb$ca$ &  $\cca$ &  \cref{eq:acti:kahn}\\
      \verb$na$ &  $\na$ &  \cref{sec:ren}\\
      \verb$Lmut$ &  $\lmut$ &  \cref{eq:abed}\\
      \verb$as$ &$\apis$ &  \cref{eq:ilka}\\
      \verb$al$ &$\apil$ &  \cref{eq:ilka}\\
      \verb$nf$ &  $\nf$ &  \cref{eq:sqcd}\\
      \verb$ep$ &  $\ep$ &  \cref{sec:action}\\
    \end{tabular}
  \end{center}
  \label{tab::variables}
\end{table}

\end{appendix}

\newpage
\bibliography{paper}

\end{document}